\documentclass[11pt]{amsart}
\usepackage[T1]{fontenc}
\usepackage{lmodern}
\usepackage{amsmath}
\usepackage{amsfonts}
\usepackage{color}      
\usepackage{epsfig}
\usepackage{mathtools}
\usepackage{amsthm}
\usepackage{bigints}
\usepackage{mathrsfs}
\usepackage{fancyhdr}
\usepackage[colorlinks=true,linkcolor=blue,citecolor=red]{hyperref}

\allowdisplaybreaks
\newcommand{\III}{\mathrm{III}}
\newcommand{\IV}{\mathrm{IV}}

\newcounter{mnotecount}
\newcommand{\mnotex}[1]
{\protect{\stepcounter{mnotecount}}$^{\mbox{\footnotesize $\bullet$\themnotecount}}$ 
\marginpar{
\raggedright\tiny\em
$\!\!\!\!\!\!\,\bullet$\themnotecount: #1} }

\begin{document}

\pagenumbering{gobble}

\newtheorem{theorem}{Theorem}[section]
\newtheorem*{theorem*}{Theorem}
\newtheorem{lemma}{Lemma}[section]
\newtheorem{remark}{Remark}[section]
\newtheorem{example}{Example}[section]
\newtheorem{proposition}{Proposition}[section]
\newtheorem{definition}{Definition}[section]
\newtheorem*{proposition*}{Proposition}

\newcommand{\comment}[1]{}

\newcommand{\blankpage}{%
    \clearpage
    \thispagestyle{empty}%
    \null
    \clearpage
}

\numberwithin{equation}{section}

\title[Projective fundamental forms at de Sitter infinity]{Projective fundamental forms at de Sitter infinity}

\def\tc{$\nabla_a^{\mathcal{T}}$}
\def\ctc{$\nabla_a^{\mathcal{T^*}}$}

\author{Jaros\l aw Kopi\'nski}
\address{Center for Theoretical Physics, Polish Academy of Sciences,
Al. Lotnik\'ow 32/46,  02-668 Warsaw, Poland}
\email{jkopinski@cft.edu.pl}

\author{Tyler Valenti}
\address{Department of Physics, University of California, Davis, CA 95616, USA}
\email{tavalenti@ucdavis.edu}

\date{\today}

\begin{abstract}
The Einstein equations constrain the geometry at infinity of asymptotically de Sitter spacetimes. We formulate such constraints for smooth projective compactifications of order two in terms of higher fundamental forms of the projective boundary. The resulting relations link the boundary curvature data encoded by these forms to the asymptotic stress--energy tensor.
\end{abstract}
\maketitle
\pagenumbering{arabic}
\section{Introduction}

Spacetimes with positive cosmological constant occupy a prominent
place in mathematical relativity and cosmology
\cite{Friedrich1986,Rendall2004,Starobinsky1983}. Geometric
compactifications make their asymptotic structure accessible by
representing infinity as a boundary of an extended spacetime
\cite{ABK2015,Frauendiener2004,Penrose1963}.
The Einstein equations can then be used to investigate how the
geometry of this boundary reflects the behaviour of the physical
fields in the asymptotic region.

In the conformal approach to infinity, the physical spacetime
$(\widetilde M,\widetilde g)$ is realized as the interior of a manifold
$M$ with boundary $\Sigma$. The unphysical metric
$g=s^2\widetilde g$ is required to extend nondegenerately
to $\Sigma$ for any boundary defining function $s$. The conformal
Einstein equations provide a framework for studying such
extensions \cite{Friedrich1983,Friedrich1986}. Their boundary restrictions
relate the geometry at infinity to asymptotic matter data
\cite{ABK2015}. Extending the equations regularly to this boundary, however, requires a formulation adapted to the matter model. Regular conformal
formulations have been developed for radiation fluids
\cite{LVK2013}, dust \cite{Friedrich2017}, and massive scalar
fields under specific assumptions on the mass and
self-interaction potential \cite{Friedrich2015}.
Related questions concerning the compatibility of matter with
conformal hypersurface geometry arise in the study of isotropic
singularities \cite{AT1999,Lubbe2013,GKW2024} and conformal
matching constructions \cite{Lubbe2013,KVK2022}.

When the unphysical metric and suitably rescaled matter fields
extend smoothly to conformal infinity, asymptotic expansions
in powers of the defining function~$s$ describe their behaviour
near~$\Sigma$ \cite{FG2012,Rendall2004,Starobinsky1983}.
The expansion coefficients record the boundary values and
successive transverse derivatives of these fields.
The Einstein equations constrain these coefficients, leaving
freely specifiable boundary data that distinguish individual
solutions \cite{Friedrich1986,Rendall2004,MPSS2016}.
Conformal fundamental forms organize geometric information
associated with transverse derivatives of the unphysical metric
into invariant boundary tensors.
In the Riemannian setting, they provide obstructions to a
conformal class admitting an asymptotically Poincar\'e--Einstein
interior metric \cite{BGW2023}. For a Poincar\'e--Einstein
filling of even dimension $d$, the critical form recovers the Neumann
data at order $s^{d-1}$ in the Fefferman--Graham expansion~\cite{BGKW2026}. Conformal fundamental forms also yield relations between boundary curvatures and stress-energy tensor of asymptotically de Sitter spacetimes~\cite{GK2023, BK2025}.

Conformal rescalings preserve null geodesic paths but change the timelike and spacelike geodesics. To retain the full family of physical
geodesics in the compactification, we consider the
projective class of the physical Levi--Civita connection
$\widetilde\nabla$. Its representatives are torsion-free connections sharing the unparametrized geodesics
of $\widetilde\nabla$ \cite{CG2016a}. Projective compactification extends this class to a manifold~$M$ with boundary $\Sigma$ by requiring a suitable
representative connection to extend smoothly to the boundary.
Let $\rho$ denote a
boundary defining function in this setting, reserving $s$ for the conformal
compactification discussed above. In the projective compactification of order two considered here, the smoothly extending representative is
obtained from $\widetilde\nabla$ by the projective change
determined by $\tfrac12\,d(\log\rho)$ \cite{CG2016a}.
Under the nondegeneracy hypotheses used below, this extension
induces a conformal structure on $\Sigma$ \cite{CG2016b}.
This provides a setting in which to relate
the asymptotic geodesic structure of the physical spacetime
to the geometry of its boundary.

To formulate the Einstein equations in the setting of projective compactification, one must
also retain the metric within the projective description. The metrizability equation supplies this link: it is a linear,
projectively invariant equation whose nondegenerate solutions
determine metrics with Levi--Civita connections in the given
projective class \cite{EM2008,FG2019}.
In spacetime dimension $d$, the physical metric supplies the solution to that equation in the form
$\lambda^{ab}:=\operatorname{vol}(\widetilde g)^{2/(d+1)}
\widetilde g^{ab}$, its density-weighted inverse metric.
Its determinant $\tau:=\det\lambda$ extends to a weight 2
defining density of the boundary~$\Sigma$~\cite{CG2016b,FG2019}.

Tractor calculus supplies invariant bundles, connections and
differential operators for conformal and projective geometries~\cite{BEG1994,CG2002,CurryGover2018}.
Its boundary applications include a conformal operator calculus~\cite{GW2014,GW2021} and a calculus relating ambient projective data
to the induced conformal boundary geometry~\cite{Porath2023}.
Applied to an asymptotically normalized defining density,
projective tractor operators generate a hierarchy of
tractor-valued higher fundamental forms~\cite{Teisseire2026}.
This raises the question of how to extract invariant
two-tensors on the projective boundary $\Sigma$.

For an order-two projective compactification, the second
fundamental form of $\Sigma$ determines the induced conformal structure
\cite{CG2016b,Porath2023}. At third and fourth order, we use
the defining density and tractor operators determined by the
physical metrizability solution to obtain explicit boundary
two-tensors. These give the first curvature constraints
beyond the induced metric. In dimension four, comparison with the
conformal fifth-form constraint of \cite{GK2023}
identifies the conformal curvature represented by the
projective fourth form. Higher orders are discussed at
the tractor level in Section~\ref{sec:fundamental}.
The main boundary constraints are as follows.

\begin{theorem}\label{thm:main}
Let $(\widetilde M,\widetilde g)$ be an asymptotically de Sitter spacetime
of dimension $d\geq4$ admitting a smooth order-two projective
compactification with boundary $\Sigma$.
Suppose that
\begin{equation}
\widetilde R_{ab}
-\tfrac12\widetilde R\widetilde g_{ab}
+\Lambda\widetilde g_{ab}
=\widetilde T_{ab},
\qquad
\widetilde T_{ab}=\tau^qT_{ab},
\quad q\in\mathbb N_{\geq1},
\label{intro:matter}
\end{equation}
with $\Lambda=(d-1)(d-2)/2$,
where $\tau$ is the weight 2 defining density of $\Sigma$
determined by the physical metric and $T_{ab}$ is a smooth tensor
of projective weight $-2q$. Then the third and fourth projective fundamental forms of $\Sigma$ satisfy
\begin{equation}
\III_{ab}=0,
\qquad
\IV_{ab}=
\begin{cases}
\displaystyle
-\tfrac{4(d-7)}{d-1}T^\top_{ab}, & q=1,\\[3pt]
0, & q>1,
\end{cases}
\label{intro:constraints}
\end{equation}
 Here $\top$ denotes tangential restriction to
$\Sigma$. Tensors $\III$ and
$\IV$ are projectively invariant and have weights $0$ and $-2$,
respectively.
\end{theorem}
The full statement, with the definitions and normalizations of
these forms, is given in Theorem~\ref{thm:boundary-constraints}
in Section~\ref{sec:fundamental}. The curvature expression for $\III$ is obtained by contracting
the transverse vector to $\Sigma$ and its conormal supplied by the metrizability
data into the projective Weyl tensor and restricting to~$\Sigma$.
Under the hypotheses of Theorem~\ref{thm:boundary-constraints},
a Gauss-type identity
also expresses $\III$ as a multiple of the trace-free difference
between the intrinsic boundary Schouten tensor and the tangential
restriction of the ambient projective Schouten tensor.
The fourth form combines a tangential divergence of projective
Weyl curvature with a projective Cotton term. In four dimensions the coefficient of the Cotton tensor term vanishes. The curvature expression for~$\IV$ contains an overall factor
$d-7$. Thus in dimension seven this form
vanishes identically, and Equation~\eqref{intro:constraints}
places no restriction on $T^\top_{ab}$. However, an intermediate boundary identity directly constrains the
combination of tangential Weyl divergence and Cotton curvature
appearing in $\IV$: it is proportional to the leading tangential
stress for $q=1$ and vanishes for $q>1$.
This identity remains valid in dimension seven and supplies
the curvature--matter constraint there; see
Section~\ref{sec:fundamental}. If the physical stress--energy tensor satisfies the null energy condition near $\Sigma$, then $T^\top_{ab}=0$
on $\Sigma$ by the argument in Section~\ref{sec:energy-conditions}. Hence
Theorem~\ref{thm:main} gives $\IV_{ab}=0$ also in the
slowest decay case $q=1$.

These results are necessary conditions for an existing smooth
order-two projective compactification. The existence of the conformal extension does not, in general,
guarantee this projective extension \cite{CG2016a}. The spatially flat Chaplygin-gas cosmology of \cite{KMP2001}
provides a non-vacuum asymptotically de Sitter example. In Section~\ref{sec:efe} we show that it admits a smooth
order-two projective compactification with the stress--energy tensor falloff $q=1$.

After the projective and tractor preliminaries,
Section~\ref{sec:efe} derives the projective Einstein--matter
equation and the boundary matter relations, then presents the
Chaplygin example. Section~\ref{sec:fundamental} relates the second fundamental
form of $\Sigma$ to the induced boundary metric, constructs the third and
fourth forms, and discusses higher-order constraints.
The detailed calculations are collected in
Appendix~\ref{sec:proofs}.

\section{Preliminaries}\label{sec:prelim}
\subsection{Projective geometry and compactification}

Let $M$ be a smooth manifold of dimension $d\geq4$ with
boundary $\Sigma:=\partial M$.
Equip its interior $\widetilde M:=\operatorname{int}M$ with
a Lorentzian metric $\widetilde g$ of signature
$(-,+,\ldots,+)$, called the \emph{physical metric}.
Tildes distinguish its Levi--Civita connection and curvature.
We follow the abstract index notation of
\cite[Chapter~2]{PenroseRindler1984}, using lower-case Latin
letters for tensor indices and upper-case Latin letters
for tractor indices.
Equality on the boundary is denoted~by~$\overset{\Sigma}{=}$.

A \emph{projective structure} $\boldsymbol p$ is an equivalence class of
torsion-free connections having the same unparametrized geodesics.
For a smooth one-form $\Upsilon$ on $M$, we write
$\widehat\nabla=\nabla+\Upsilon$ for the projective change
of connection defined by
\begin{equation*}
 \widehat\nabla_a v^b=\nabla_a v^b+\Upsilon_a v^b
                     +\delta_a{}^b\Upsilon_cv^c,
 \qquad
 \widehat\nabla_a u_b=\nabla_a u_b-\Upsilon_a u_b-\Upsilon_bu_a,
\end{equation*}
for $v\in\Gamma(TM)$ and $u\in\Gamma(T^*M)$.
The bundle of projective densities of weight $w$ is $\mathcal E(w):=((\Lambda^dTM)^{\otimes2})^{w/(2d+2)}$.
For any tensor bundle $\mathcal V$ over $M$, we write
$\mathcal V(w):=\mathcal V\otimes\mathcal E(w)$.
In abstract-index notation, $\mathcal E^a:=TM$ and
$\mathcal E_a:=T^*M$.
For $f\in\Gamma(\mathcal E(w))$, the induced connections satisfy
\begin{equation*}
 \widehat\nabla_a f=\nabla_af+w\Upsilon_af.
\end{equation*}
A positive section $\varphi \in\Gamma(\mathcal E(1))$ determines a
unique representative $\nabla\in\boldsymbol p$ satisfying
$\nabla_a\varphi=0$~\cite[Section~2.2]{CG2016a}.
Such a representative is called \emph{special}: it preserves a
nowhere-vanishing volume density. We use special representatives of $\boldsymbol p$
for the calculations below. Their Ricci and Schouten tensors are symmetric, and
projective changes between these representatives are
determined by locally exact one-forms $\Upsilon$
\cite{EM2008}. Our curvature conventions are 
\begin{equation*}
 [\nabla_a,\nabla_b]v^c=:R_{ab}{}^c{}_ev^e,
 \qquad
 R_{ab}{}^c{}_e=W_{ab}{}^c{}_e+2\delta_{[a}{}^cP_{b]e},
 \qquad R_{ab}:=R_{ca}{}^c{}_b, \qquad 
 P_{ab}:=\tfrac{R_{ab}}{d-1}.
\end{equation*}
Here $R_{ab}$ is a projective Ricci tensor, $P_{ab}$ is a projective Schouten tensor and $W_{ab}{}^c{}_e$ is a trace-free projective
Weyl tensor. The Cotton tensor can be defined as
\begin{equation*}
 C_{abc}:=2\nabla_{[a}P_{b]c},
\end{equation*}
and the contracted differential Bianchi identity gives
\begin{equation}\label{pre:contracted-bianchi}
 \nabla_cW_{ab}{}^c{}_e=(d-2)C_{abe}.
\end{equation}
 Under a projective change,
\begin{equation*}
 \widehat P_{ab}=P_{ab}-\nabla_a\Upsilon_b+\Upsilon_a\Upsilon_b,
 \qquad \widehat W_{ab}{}^c{}_e=W_{ab}{}^c{}_e,
 \qquad \widehat C_{abc}=C_{abc}+\Upsilon_eW_{ab}{}^e{}_c.
\end{equation*}

A \emph{defining function} $\rho$ of $\Sigma$ satisfies $\Sigma=\rho^{-1}(0)$ and $d \rho|_\Sigma\neq0$.
The physical metric $\widetilde g$ is \emph{projectively compact of order two} if
\begin{equation*}
 \nabla^{(\rho)}:=\widetilde\nabla+\tfrac{d \rho}{2 \rho}
\end{equation*}
extends smoothly to $M$. This condition is independent of $\rho$ and
extends the physical projective structure to $M$. A section $\tau\in\Gamma(\mathcal E(2))$ is a \emph{defining density} for $\Sigma$ if it can be written as
$\tau=\rho\eta$, where $\eta$ is a smooth nowhere-vanishing weight 2 density.
We write $U=O(\tau^k)$ when $U=\tau^kV$ with~$V$ smooth and of the
appropriate weight.

\subsection{Tractors and the Thomas operator}

We recall the projective tractor calculus needed below; detailed
accounts are given in \cite{BEG1994,CG2002} and
\cite[Section~3.1]{CG2016a}.
The standard cotractor bundle is $\mathcal T^*:=J^1\mathcal E(1)$,
with dual $\mathcal T$. Here $J^1\mathcal E(1)$ is the first jet
bundle: its fibre at a point records the value and first derivative
of a local section of $\mathcal E(1)$ at that point. It fits into
the exact sequence
\begin{equation*}
 0\longrightarrow\mathcal E_a(1)
 \longrightarrow\mathcal T^*
 \longrightarrow\mathcal E(1)\longrightarrow0.
\end{equation*}
A representative connection $\nabla\in\boldsymbol p$ identifies
this sequence with the direct sum
$\mathcal E(1)\oplus\mathcal E_a(1)$. This identification is called
a \emph{splitting}. Using the splitting and its dual, sections $U_A\in\Gamma(\mathcal T^*)$ and
$V^A\in\Gamma(\mathcal T)$ can be written as
\begin{equation*}
 U_A=Y_Au+Z_A{}^au_a,
 \qquad
 V^A=W_a{}^Av^a+X^Av.
\end{equation*}
Here $X^A$ is the canonical projection $U_A\mapsto u=X^AU_A$,
and $Z_A{}^a$ is the canonical inclusion
$u_a\mapsto Z_A{}^au_a$ in the jet sequence.
The complementary maps $Y_A$ and $W_a{}^A$ depend on $\nabla$;
they give $u_a=W_a{}^AU_A$ and $v=Y_AV^A$. The weights of $X^A,Z_A{}^a,Y_A,W_a{}^A$
are $1,-1,-1$, and $1$, respectively.
The components of a tractor in a splitting will sometimes be called \emph{slots}. For the first jet of a section $f\in\Gamma(\mathcal E(1))$, the slots are $(f,\nabla_af)$. The splitting maps satisfy
\begin{equation*}
 \begin{gathered}
 X^AY_A=1,\qquad X^AZ_A{}^a=0,\qquad W_a{}^AY_A=0,
 \qquad W_a{}^AZ_A{}^b=\delta_a{}^b,\\
 \delta_A{}^B=Y_AX^B+Z_A{}^aW_a{}^B.
 \end{gathered}
\end{equation*}
Under a projective change $\widehat\nabla=\nabla+\Upsilon$, they
transform as
\begin{equation}\label{pre:splitting-change}
 \begin{aligned}
 \widehat X^A&=X^A,
 &\widehat Z_A{}^a&=Z_A{}^a,\\
 \widehat Y_A&=Y_A-\Upsilon_aZ_A{}^a,
 &\widehat W_a{}^A&=W_a{}^A+\Upsilon_aX^A.
 \end{aligned}
\end{equation}

The projective structure determines a canonical tractor connection
on $\mathcal T$ and a dual connection on $\mathcal T^*$. These are independent of the
representative $\nabla$ used to express them. In a chosen
splitting, the columns $\binom{v^b}{v}$ and $\binom{u}{u_b}$
denote the ordered slots of $V^A$ and $U_A$, respectively.
The connections act by
\begin{equation*}
 \nabla_a^{\mathcal T}\begin{pmatrix}v^b\\v\end{pmatrix}
 =\begin{pmatrix}\nabla_av^b+\delta_a{}^bv\\
                  \nabla_av-P_{ab}v^b\end{pmatrix},
 \qquad
 \nabla_a^{\mathcal T^*}\begin{pmatrix}u\\u_b\end{pmatrix}
 =\begin{pmatrix}\nabla_au-u_a\\
                  \nabla_au_b+P_{ab}u\end{pmatrix}.
\end{equation*}
The derivatives on the right are induced by $\nabla$ on the
appropriate tensor and density bundles. Henceforth we suppress
the superscripts $\mathcal T$ and $\mathcal T^*$ and also use
$\nabla$ for the induced connections on their tensor products
and density-weighted versions. In particular,
\begin{equation*}
 \nabla_aX^B=W_a{}^B,\qquad
 \nabla_aW_b{}^B=-P_{ab}X^B,\qquad
 \nabla_aY_B=P_{ab}Z_B{}^b,\qquad
 \nabla_aZ_B{}^b=-\delta_a{}^bY_B.
\end{equation*}
The tractor curvature $\kappa_{ab}{}^C{}_D$ is defined by
$[\nabla_a,\nabla_b]V^C=:\kappa_{ab}{}^C{}_DV^D$.
In terms of the projective Weyl and Cotton tensors, it reads
\begin{equation*}
 \kappa_{ab}{}^C{}_D
 =W_{ab}{}^c{}_eW_c{}^CZ_D{}^e-C_{abc}X^CZ_D{}^c.
\end{equation*}

For a collection $\Phi$ of tractor and cotractor indices, let
$\mathcal E^\Phi(w)$ denote the corresponding tractor tensor
bundle of projective weight $w$. The projective Thomas operator
is the canonical first-order, projectively invariant differential
operator \cite{BEG1994,FG2019}, with domain and range
\begin{equation*}
 D_A:\Gamma(\mathcal E^\Phi(w))\longrightarrow
       \Gamma(\mathcal E_A{}^\Phi(w-1)).
\end{equation*}
It adds a cotractor index and lowers the
weight by one. In a splitting, its action on any section
$F\in\Gamma(\mathcal E^\Phi(w))$ is
\begin{equation*}
 D_AF=wY_AF+Z_A{}^a\nabla_aF.
\end{equation*}
The operator
obeys the Leibniz rule and satisfies
\begin{equation*}
 X^AD_AF=wF,\qquad D_AX^B=\delta_A{}^B,\qquad
 D_{[A}D_{B]}f=0,
\end{equation*}
for $f\in\Gamma(\mathcal E(w))$. Its commutator on tractor fields is determined by $\kappa$.
For~$F^C\in\Gamma(\mathcal T(w))$,
\begin{equation*}
 [D_A,D_B]F^C
 =Z_A{}^aZ_B{}^b\kappa_{ab}{}^C{}_DF^D.
\end{equation*}
On a field with several tractor indices, the curvature acts on
each contravariant index in this way and with the dual  action on each covariant index.

\subsection{Metrizability and boundary data}

The physical metric determines the symmetric weight~$-2$ tensor
$\lambda^{ab}:=\operatorname{vol}(\widetilde g)^{2/(d+1)}
\widetilde g^{ab}$, where $\operatorname{vol}(\widetilde g)$ denotes
the metric volume density. It satisfies the projectively invariant
metrizability equation
\begin{equation}\label{pre:metrizability}
 \nabla_a\lambda^{bc}=-\delta_a{}^b\nu^c-\delta_a{}^c\nu^b,
 \qquad \nu^b:=-\tfrac1{d+1}\nabla_c\lambda^{bc}.
\end{equation}
We will call $\lambda$ the \emph{physical metrizability solution}. Taking the determinant of $\lambda$ using the canonical weighted volume gives
a density $\tau$ of weight $2$,
\begin{equation}\label{pre:tau-determinant}
 \tau:=\det\lambda=-\operatorname{vol}(\widetilde g)^{-2/(d+1)},
 \qquad \widetilde g^{ab}=|\tau|\lambda^{ab}.
\end{equation}
For an order-two projective compactification, $\tau$ extends to
a defining density for $\Sigma$, and $\lambda$ extends smoothly
with rank $d-1$ on $\Sigma$ \cite{CG2016b}.
Metric compatibility gives $\widetilde\nabla\tau=0$ on
$\widetilde M$. Thus fixing $\widetilde g$ fixes both the physical metrizability solution $\lambda$ and its determinant $\tau$.

The projectively invariant splitting operator $L$ lifts a
symmetric weight $-2$ tensor to a symmetric tractor whose
projection onto $S^2TM(-2)$ is the original tensor
\cite[Proposition~3.1]{CGM2014}. Its additional slots involve
first and second derivatives. Applied to $\lambda$, it gives
the smooth weight 0 tractor
\begin{equation*}
 H^{AB}:=L(\lambda)^{AB}
 =\lambda^{ab}W_a{}^AW_b{}^B
   +2\nu^aW_a{}^{(A}X^{B)}+\sigma X^AX^B,
\end{equation*}
where $\sigma:=\tfrac1d(P_{ab}\lambda^{ab}-\nabla_a\nu^a)$. Differentiating the metrizability equation and commuting
covariant derivatives yields the equations for the additional
slots of $H$,
\begin{align}
 \nabla_a\nu^b
 &=P_{ac}\lambda^{bc}-\delta_a{}^b\sigma
   -\tfrac1dW_{ac}{}^b{}_e\lambda^{ce},
 \label{eq:prolongation-middle}\\
 \nabla_a\sigma
 &=2P_{ab}\nu^b+\tfrac2dC_{abc}\lambda^{bc}.
 \label{pre:prolongation-bottom}
\end{align}
Together with Equation~\eqref{pre:metrizability}, these form a
closed first-order system for $(\lambda^{ab},\nu^a,\sigma)$,
called the prolongation of the metrizability equation
\cite[Theorem~4.1]{CGM2014}. The trace-free property of the Weyl
tensor then gives $D_AH^{AB}=0$.
The stronger condition $\nabla_aH^{BC}=0$, with respect
to the canonical tractor connection, is equivalent to
$\widetilde g$ being Einstein on $\widetilde M$
\cite[Theorem~3.3]{CGM2014}.

The canonical parallel tractor volume squared defines
determinants of symmetric tractors. Locally, choose a
normalized tractor volume form
$\epsilon_{A_0\ldots A_d}$ and write
\cite[Section~2.1]{FG2019}
\begin{equation*}
 \alpha:=\det(H^{AB})
 =\tfrac1{(d+1)!}
   \epsilon_{A_0\ldots A_d}\epsilon_{B_0\ldots B_d}
   \prod_{i=0}^{d}H^{A_iB_i}.
\end{equation*}
On $\widetilde M$, in the physical Levi--Civita splitting,
$\nu^a=0$ and
$\alpha=\tau\sigma=-\widetilde R/(d(d-1))$. For an order-two projectively compact metric, the physical scalar
curvature extends smoothly to the boundary with a locally constant,
nonzero boundary value \cite[Theorem~7]{CG2016b}.
Consequently, $\alpha$ is nonzero near~$\Sigma$ and locally constant
along~$\Sigma$, and $H^{AB}$ is nondegenerate there.
Let $H_{AB}$ be the inverse of $H^{AB}$, characterized by
$H_{AC}H^{CB}=\delta_A{}^B$, and write
\begin{equation*}
 H_{AB}:=\beta Y_AY_B+2\mu_aZ_{(A}{}^aY_{B)}
                         +\zeta_{ab}Z_A{}^aZ_B{}^b.
\end{equation*}
The coefficients $\beta,\mu_a,\zeta_{ab}$ have weight $2$. Cofactor identities for $H^{AB}$, together with the metrizability
equation, yield the identities
\cite[Corollary~5.4.2, Lemma~5.4.4 and Proposition~5.4.5]{Porath2023}
\begin{equation}\label{pre:determinant-identities}
 \tau=\alpha\beta,\qquad D_A\tau=2\alpha H_{AB}X^B,\qquad
 \nabla_a\tau=2\alpha\mu_a.
\end{equation}
Expanding the inverse relation $H_{AC}H^{CB}=\delta_A{}^B$ also gives
\begin{equation}\label{pre:inverse-slots}
 \begin{gathered}
 \lambda^{ab}\mu_b=-\beta\nu^a,\qquad
 \mu_a\nu^a=1-\beta\sigma,\\
 \zeta_{ab}\lambda^{bc}=\delta_a{}^c-\mu_a\nu^c,\qquad
 \zeta_{ab}\nu^b=-\sigma\mu_a.
 \end{gathered}
\end{equation}

On $\Sigma$, the identity $\tau=\alpha\beta$ and
$\alpha\neq0$ imply $\beta=0$. Hence the preceding relations
reduce to $\mu_a\nu^a=1$ and $\lambda^{ab}\mu_b=0$ there.
Moreover, $\nabla_a\tau=2\alpha\mu_a$ shows that $\mu_a$ is a
nonvanishing weighted conormal, so $\nu^a$ is transverse to
$\Sigma$. In a neighbourhood of $\Sigma$, define the smooth ambient
endomorphisms
\begin{equation}\label{pre:boundary-projectors}
 \Pi_a^{\top b}:=\delta_a{}^b-\mu_a\nu^b,
 \qquad
 \Pi_a^{\perp b}:=\mu_a\nu^b.
\end{equation}
On $\Sigma$, the identity $\mu_a\nu^a=1$ makes
$\Pi^\top$ and $\Pi^\perp$ complementary projections.
On covectors, $\Pi^\perp$ projects onto the conormal line
spanned by $\mu$, while $\Pi^\top$ projects onto covectors
annihilating $\nu$. Restriction identifies the latter space
with $T^*\Sigma$. The dual of $\Pi^\top$ projects vectors
onto $T\Sigma$ along the transverse direction determined
by $\nu$. For a covariant two-tensor $t_{ab}$, set
$$
 t^\top_{ab}:=\Pi_a^{\top c}\Pi_b^{\top e}t_{ce}.
$$
Along $\Sigma$, a change of splitting gives
$\widehat\mu_a=\mu_a$ and
$\widehat\nu^a=\nu^a-\lambda^{ab}\Upsilon_b$.
The relation $\lambda^{ab}\mu_b=0$ shows that the change
in $\nu^a$ is tangential.
Since $\operatorname{rank}(\lambda)\overset{\Sigma}{=}d-1$, the kernel of
$\lambda$ is precisely the conormal line spanned by~$\mu$.
For covector fields $u,v\in\Gamma(\mathcal E_a)$,
the contraction $\lambda^{ab}u_av_b\big|_\Sigma$
is unchanged when a conormal covector is added to either
argument.
Denote the induced nondegenerate weighted symmetric bilinear
form on $T^*\Sigma$ by $\gamma^{ab}$ and its inverse by
$\gamma_{ab}$.
Their projective weights are $-2$ and $2$, respectively.
Choosing a density scale turns $\gamma_{ab}$ into an
ordinary boundary metric representing the induced conformal
structure~\cite{FG2019}.
Bars denote the intrinsic Levi--Civita connection and
curvature of a chosen boundary metric.
\subsection{The tractor Laplacian and extraction}

The physical metrizability solution determines the Thomas Laplacian
\begin{equation*}
 \Box:\Gamma(\mathcal E^\Phi(w))\longrightarrow
       \Gamma(\mathcal E^\Phi(w-2)),
 \qquad \Box V:=H^{AB}D_AD_BV.
\end{equation*}
In a splitting,
\begin{equation*}
 \Box V=\lambda^{ab}\nabla_a\nabla_bV
       +2(w-1)\nu^a\nabla_aV
       +w\big(P_{ab}\lambda^{ab}+(w-1)\sigma\big)V.
\end{equation*}
To determine how $\Box$ acts on products of powers of the defining density $\tau$
with smooth tractor fields, we use the commutator identity
\begin{equation}\label{pre:box-commutator}
 [\Box,\tau]V=2\alpha(d+2w+1)V,
\end{equation}
which follows from Equation~\eqref{pre:determinant-identities} and
$D_AH^{AB}=0$ \cite[Theorem~7.1.9]{Porath2023}. In particular, if $V=\tau^k U$ with $U$ smooth and $k\geq1$, then $\Box V=O(\tau^{k-1})$.

Let $T_{A_1\ldots A_r}$ be a covariant tractor of weight $w$
whose contraction with $X$ in each slot vanishes. Thus, in any splitting, the only component of $T$ that can be nonzero is the one with a factor of $Z$ in every slot. There is therefore a unique
$t\in\Gamma(\mathcal E_{a_1\ldots a_r}(w+r))$ such that
\begin{equation}\label{pre:pure-slot-extraction}
 T_{A_1\ldots A_r}
 =Z_{A_1}{}^{a_1}\cdots Z_{A_r}{}^{a_r}t_{a_1\ldots a_r}.
\end{equation}
\emph{Extraction} means recovering the weighted tensor $t$ from
$T$ by inverting this inclusion on its image. We denote
the resulting tensor by $p^*(T):=t$. Since $Z$ is projectively
invariant by Equation~\eqref{pre:splitting-change}, the unique
tensor $t$ is independent of the splitting. When terms containing $Y$ occur in the constructions below, we first remove them by differential corrections and then
apply $p^*$.
\section{The Projective Einstein-Matter Equation and its consequences}\label{sec:efe}
Assume that $\widetilde M$ is connected, $\Sigma\ne\emptyset$,
and the physical metric $\widetilde g$ is projectively compact
of order two. We consider the Einstein field equations with
cosmological constant $\Lambda$,
\begin{equation}\label{eq:efe}
 \widetilde R_{ab}-\tfrac12\widetilde R\widetilde g_{ab}
 +\Lambda\widetilde g_{ab}=\widetilde T_{ab}.
\end{equation}
Since $\widetilde R_{ab}=(d-1)\widetilde P_{ab}$, their
trace-reversed form is
\begin{equation}\label{eq:trace-reversed}
 \widetilde P_{ab}=c_\Lambda\widetilde g_{ab}+\widetilde S_{ab},
 \qquad
 \widetilde S_{ab}:=\tfrac1{d-1}
 \left(\widetilde T_{ab}
       -\tfrac{\widetilde T}{d-2}\widetilde g_{ab}\right),
 \qquad
 c_\Lambda:=\tfrac{2\Lambda}{(d-1)(d-2)},
\end{equation}
where $\widetilde T=\widetilde g^{ab}\widetilde T_{ab}$.
Since $c_\Lambda$ is constant and
$\widetilde\nabla\widetilde g=0$, differentiation gives
\begin{equation}\label{eq:differentiated-efe}
 \widetilde\nabla_c(\widetilde P_{ab}-\widetilde S_{ab})=0.
\end{equation}
This equation serves as the starting point for our projective formulation. It states that the tensor $\widetilde P-\widetilde S$ is parallel with respect to the physical Levi--Civita connection. In the setting considered here, every such tensor is a constant multiple of the metric,
\begin{equation}\label{eq:parallel-rigidity}
 \big\{K\in\Gamma(S^2T^*\widetilde M):\widetilde\nabla K=0\big\}
 =\operatorname{span}_{\mathbb R}\{\widetilde g\}.
\end{equation}
This is proved in Appendix~\ref{app:efe-equivalence}. It follows that
\begin{equation}\label{eq:efe-equivalence}
 \widetilde\nabla_c(\widetilde P_{ab}-\widetilde S_{ab})=0
 \quad\Longleftrightarrow\quad
 \widetilde P_{ab}-\widetilde S_{ab}=c_0\widetilde g_{ab}
 \quad\text{for some }c_0\in\mathbb R.
\end{equation}
Hence we see that Equation~\eqref{eq:differentiated-efe} is equivalent to the trace-reversed Einstein equation with cosmological constant
$\tfrac12(d-1)(d-2)c_0$, determined by
the given metric and stress--energy tensor. For a solution of
Equation~\eqref{eq:efe}, this recovers the prescribed value
of the cosmological constant $\Lambda$. In the boundary calculations below,  we take solutions of
Equation~\eqref{eq:efe} with $\Lambda=(d-1)(d-2)/2$. Then $c_0=c_\Lambda=1$ and
$\widetilde P_{ab}=\widetilde g_{ab}+\widetilde S_{ab}$.

Let $\nabla$ be a smooth special connection in the projective class $\boldsymbol p$ on $M$ supplied by the order-two
compactification. Since $\boldsymbol p|_{\widetilde M}=[\widetilde\nabla]$, we can write
$\widetilde\nabla=\nabla+\Upsilon$ on $\widetilde M$. Denote the projective Schouten tensor of
$\nabla$ by $P_{ab}$. Recall that $\tau=\det\lambda$ is the
weight 2 defining density of $\Sigma$ obtained from the
weighted inverse metric $\lambda^{ab}$. This density is parallel for
the physical Levi--Civita connection $\widetilde\nabla$. The projective change therefore gives
\begin{equation} \label{eq:ptop}
 \widetilde P_{ab}
 =P_{ab}-\nabla_a\Upsilon_b+\Upsilon_a\Upsilon_b,
 \qquad
 0=\widetilde\nabla_a\tau=\nabla_a\tau+2\tau\Upsilon_a.
\end{equation}
On the interior, where $\tau\neq0$, the second identity gives
$\Upsilon_a=-\nabla_a\tau/(2\tau)$. We now want to use Equation~\eqref{eq:ptop} to write Equation~\eqref{eq:differentiated-efe} in terms of the connection $\nabla$, its projective Schouten tensor~$P$ and weight 2 defining density $\tau$. This yields
\begin{equation}\label{eq:omega-efe}
\Omega_{abc} =  2\tau\widetilde\nabla_c\widetilde S_{ab},
\end{equation}
where
\begin{equation}\label{eq:omega}
 \Omega_{abc}:=\nabla_c\nabla_a\nabla_b\tau
   +2P_{ab}\nabla_c\tau
   +2\tau\nabla_cP_{ab}
   +2P_{c(b}\nabla_{a)}\tau.
\end{equation}
The tensor $\Omega_{abc}$ is smooth up to $\Sigma$ because
$\nabla$, $P_{ab}$ and $\tau$ are smooth. To obtain its tractor
expression, set $I_A:=D_A\tau$. The splitting rules give
\[
 D_AI_B
 =2\tau Y_AY_B+2\nabla_a\tau Z_{(A}{}^aY_{B)}
  +(\nabla_a\nabla_b\tau+2\tau P_{ab})Z_A{}^aZ_B{}^b.
\]
This cotractor has weight zero, so
$D_C=Z_C{}^c\nabla_c$ when acting on it. Differentiating the splitting
maps cancels all terms containing $Y_A$ or $Y_B$, leaving
\begin{equation}\label{eq:dddtau}
 D_CD_AI_B=Z_C{}^cZ_A{}^aZ_B{}^b\Omega_{abc}.
\end{equation}
Equation~\eqref{eq:dddtau} identifies $\Omega_{abc}$ as the
tensor obtained by the algebraic extraction $p^*(DDI)$. Since the tractor $DDI$
and the inclusion map are projectively invariant, so is
$\Omega$.

In the physical splitting, $\widetilde\nabla\tau=0$, so
$\Omega_{abc}=2\tau\widetilde\nabla_c\widetilde P_{ab}$.
Equation~\eqref{eq:differentiated-efe} therefore gives
Equation~\eqref{eq:omega-efe}. Up to normalization,
$\Omega_{(abc)}$ is the first projective Bernstein–Gelfand–Gelfand (BGG) operator on
weight 2 densities applied to $\tau$~\cite{CG2016a}.

We now turn to the stress--energy tensor and consider the following assumption about the decay of matter fields as we approach the projective boundary $\Sigma$,
\begin{equation*}
\widetilde T_{ab}=\tau^qT_{ab},\qquad q\in\mathbb N_{\geq1},
\end{equation*}
where the \emph{unphysical stress--energy tensor} $T_{ab}$ is smooth up to $\Sigma$ and has weight $-2q$. Define its weighted trace as $\theta:=\lambda^{ab}T_{ab}$. We will also make use of the \emph{unphysical trace-corrected stress--energy tensor} that can be defined as
\[
 S_{ab}:=\tau^{-q}\widetilde S_{ab}
 =\tfrac{1}{d-1}\left(
 T_{ab}-\tfrac{\theta}{d-2}(\lambda^{-1})_{ab}
 \right).
\]
Since $\widetilde\nabla\tau=0$, Equation~\eqref{eq:omega-efe} now becomes
\begin{equation}\label{eq:omega-matter}
 \Omega_{abc}=2\tau^{q+1}\widetilde\nabla_c S_{ab}.
\end{equation}
To compute the boundary value of $\Omega_{abc}$ we use stress--energy conservation. Taking the divergence
of Equation~\eqref{eq:efe} with $\widetilde\nabla$ and using the contracted
Bianchi identity for $\widetilde g$ gives
\begin{equation}\label{eq:physical-conservation}
 0=\widetilde g^{ac}\widetilde\nabla_c\widetilde T_{ab}
   =|\tau|\tau^q\lambda^{ac}\widetilde\nabla_cT_{ab},
\end{equation}
where the second equality uses Equation~\eqref{pre:tau-determinant} and
$\widetilde\nabla\tau=0$. Thus
$\lambda^{ac}\widetilde\nabla_cT_{ab}=0$ on the interior.
For the weight $-2q$ tensor $T_{ab}$, the projective change gives
\[
 \widetilde\nabla_cT_{ab}
 =\nabla_cT_{ab}-2(q+1)\Upsilon_cT_{ab}
               -\Upsilon_aT_{cb}-\Upsilon_bT_{ac},
 \qquad \Upsilon_a=-\tfrac{\alpha}{\tau}\mu_a.
\]
Contracting with $\lambda^{ac}$ and multiplying by $\tau$ yields
\begin{equation}\label{eq:trace-conservation-proof}
 0=\tau\lambda^{ac}\nabla_cT_{ab}
   +(2q+3)\alpha\lambda^{ac}\mu_cT_{ab}+\alpha\mu_b\theta .
\end{equation}
The first two terms vanish on $\Sigma$ because
$\tau\overset{\Sigma}{=}0$ and $\lambda^{ab}\mu_b\overset{\Sigma}{=}0$.
Since $\alpha\mu_b$ is nonzero there, for every~$q\geq1$ we obtain
\begin{equation}\label{eq:trace-zero}
 \theta=\lambda^{ab}T_{ab}\overset{\Sigma}{=}0.
\end{equation}
Thus $\theta=O(\tau)$ and $\theta/\tau$ extends smoothly to $\Sigma$. The relations $\tau=\alpha\beta$ and $\nabla_a \tau=2\alpha\mu_a$
from Equation~\eqref{pre:determinant-identities}, together with Equation~\eqref{pre:inverse-slots}, imply
\begin{equation}\label{eq:matter-inverse}
 \lambda^{ab}\nabla_b\tau=-2\tau\nu^a,\qquad
 (\lambda^{-1})_{ab}=\zeta_{ab}-\tfrac{\alpha}{\tau}\mu_a\mu_b.
\end{equation}
For the second identity, multiplying its right-hand side by
$\lambda^{bc}$ gives $\delta_a{}^c$ by Equation~\eqref{pre:inverse-slots}. By Equation~\eqref{eq:matter-inverse}, the product
$\theta(\lambda^{-1})_{ab}
=\theta\zeta_{ab}-\tfrac{\alpha\theta}{\tau}\mu_a\mu_b$
extends smoothly to $\Sigma$. Together with the assumed smoothness of $T_{ab}$, this
proves that $S_{ab}$ extends smoothly to $\Sigma$.

Set $S^\top_{ab}:=\Pi_a^{\top c}\Pi_b^{\top e}S_{ce}$.
On $\Sigma$, $\Pi^\top$ annihilates $\mu_a$, and
$\lambda^{ab}T^\top_{ab}=\theta$. Equation~\eqref{eq:trace-zero}
therefore gives
\[
 (d-1)S^\top_{ab}\overset{\Sigma}{=}T^\top_{ab},
 \qquad
 \lambda^{ab}S^\top_{ab}\overset{\Sigma}{=}0.
\]
For the weight $-2q$ tensor $S_{ab}$, the projective change
of connection gives
\begin{align}
 2\tau^{q+1}\widetilde\nabla_c S_{ab}
 &=2\tau^{q+1}\nabla_c S_{ab}
 +2\alpha\tau^q\bigl[2(q+1)\mu_c S_{ab}
 +\mu_a S_{cb}+\mu_b S_{ac}\bigr].
 \label{eq:matter-change}
\end{align}
Each term on the right-hand side is $O(\tau^q)$ because
$S_{ab}$ and $\nabla_cS_{ab}$ are smooth up to $\Sigma$.
Equation~\eqref{eq:omega-matter} therefore gives
$\Omega_{abc}=O(\tau^q)$. Since $q\geq1$, we obtain
\begin{equation}\label{eq:omega-zero}
 \Omega_{abc}\overset{\Sigma}{=}0.
\end{equation}

We next show that the stress--energy falloff
$\widetilde T_{ab}=\tau^qT_{ab}$ forces the Weyl and Cotton
contractions in Equations~\eqref{eq:prolongation-middle}
and~\eqref{pre:prolongation-bottom} to vanish on $\Sigma$. In the physical splitting
$\widetilde\nabla_a\lambda^{bc}=0$, so $\nu^a=0$ by
Equation~\eqref{pre:metrizability}. The bottom slot $\sigma$ of~$H^{AB}$
then has the value $\sigma=\tfrac1d\widetilde P_{ac}\lambda^{ac}$
by its definition in Section~\ref{sec:prelim}. Equation~\eqref{eq:prolongation-middle} therefore gives
\[
 W_{ac}{}^b{}_e\lambda^{ce}
 =d\left(\widetilde P_{ac}\lambda^{bc}
 -\tfrac1d\widetilde P_{ce}\lambda^{ce}\delta_a{}^b\right).
\]
Since $\widetilde g_{ac}\lambda^{bc}=|\tau|^{-1}\delta_a{}^b$,
the cosmological term and the correction proportional to
$\widetilde T\widetilde g_{ac}$ in Equation~\eqref{eq:trace-reversed}
each give a multiple of $\delta_a{}^b$ upon contraction with
$\lambda^{bc}$. Both contributions therefore vanish in the
trace-free combination above. Substituting
$\widetilde T_{ab}=\tau^qT_{ab}$ gives
\begin{equation}\label{eq:Wlambda-matter}
 W_{ac}{}^b{}_e\lambda^{ce}
 =\tfrac{d}{d-1}\tau^q\left(\lambda^{bc}T_{ac}
                 -\tfrac1d\theta\delta_a{}^b\right).
\end{equation}
For the contraction of the Cotton tensor with $\lambda$, use the
contracted projective Bianchi identity
\eqref{pre:contracted-bianchi} in the smooth splitting.
Using the product rule, Equation~\eqref{pre:metrizability} gives
\[
 W_{ac}{}^b{}_e\nabla_b\lambda^{ce}
 =-W_{ac}{}^c{}_e\nu^e-W_{ac}{}^b{}_b\nu^c=0
\]
by trace-freeness of $W$. Consequently,
\begin{equation}\label{eq:div-Wlambda}
 \nabla_b(W_{ac}{}^b{}_e\lambda^{ce})=(d-2)C_{ace}\lambda^{ce},
\end{equation}
and Equation~\eqref{eq:Wlambda-matter} yields
\begin{equation}\label{eq:C-matter}
 C_{abc}\lambda^{bc}
 =\tfrac{d}{(d-1)(d-2)}\nabla_b
 \left[\tau^q\left(\lambda^{bc}T_{ac}
                   -\tfrac1d\theta\delta_a{}^b\right)\right].
\end{equation}
The term on the right-hand side where the derivative acts on $\tau^q$ is proportional to $\tau^{q-1}$ times
\[
 \lambda^{bc}(\nabla_b\tau)T_{ac}-\tfrac1d\theta\nabla_a\tau
 =-2\tau\nu^bT_{ab}-\tfrac1d\theta\nabla_a\tau=O(\tau).
\]
The other terms already have a factor of $\tau^q$. Thus,
\[
 W_{ac}{}^b{}_e\lambda^{ce}=O(\tau^q),\qquad
 C_{abc}\lambda^{bc}=O(\tau^q).
\]
Using the tractor connection formulas and Equations~\eqref{pre:metrizability}, \eqref{eq:prolongation-middle} and
\eqref{pre:prolongation-bottom}, we have
\[
 \nabla_aH^{BC}
 =-\tfrac2d W_{ac}{}^b{}_e\lambda^{ce}W_b{}^{(B}X^{C)}
  +\tfrac2d C_{abc}\lambda^{bc}X^BX^C.
\]
Hence $\nabla_aH^{BC}=O(\tau^q)$. In particular, the
\emph{boundary normality condition} holds:
\begin{equation}\label{eq:normality}
\nabla_aH^{BC}\overset{\Sigma}{=}0.
\end{equation}
The display above implies that the inverse tractor satisfies
$\nabla_aH_{BC}\overset{\Sigma}{=}0$. The mixed slot of this
derivative is~$\nabla_a\mu_b+\beta P_{ab}-\zeta_{ab}$.
Since $\beta\overset{\Sigma}{=}0$, we obtain
\[
 \nabla_a\mu_b\overset{\Sigma}{=}\zeta_{ab}.
\]

The curvature formulas needed later also involve
the boundary value of the tractor determinant~$\alpha$. Since
$\alpha=-\widetilde R/(d(d-1))$, the trace of the Einstein equation
determines this value and the order to which $\alpha$ approaches it.
Equation~\eqref{eq:trace-zero} gives $\theta=O(\tau)$. Using
$\widetilde g^{ab}=|\tau|\lambda^{ab}$ and $\tau<0$ from Equation~\eqref{pre:tau-determinant}, we obtain
\[
 \widetilde T=\widetilde g^{ab}\widetilde T_{ab}
 =|\tau|\tau^q\theta=-\tau^{q+1}\theta=O(\tau^{q+2}).
\]
Since $\Lambda=(d-1)(d-2)/2$, tracing Equation~\eqref{eq:efe} gives
$\widetilde R=d(d-1)-2\widetilde T/(d-2)$. Consequently,
\[
 \alpha=-1-\tfrac{2\tau^{q+1}\theta}{d(d-1)(d-2)}
       =-1+O(\tau^{q+2}).
\]
In particular, $\alpha\overset{\Sigma}{=}-1$ and
$\nabla_a\alpha\overset{\Sigma}{=}0$.

We conclude this section with an explicit non-vacuum
spacetime admitting a smooth projective compactification
of order two.
\begin{example}\label{ex:chaplygin}
For the spatially flat four-dimensional Chaplygin-gas
cosmology of \cite{KMP2001}, let $a$ denote the
Friedmann–Lemaître–Robertson–Walker scale factor and
put $\rho=a^{-2}$. The metric is
\begin{equation}\label{eq:chaplygin-metric}
 \widetilde g=-\tfrac{d\rho^2}{4\rho^2F(\rho)}
              +\rho^{-1}\delta_{ij}dx^idx^j,
 \qquad F(\rho)=\sqrt{1+b\rho^3},\qquad b>0.
\end{equation}
For $\nabla=\widetilde\nabla+d\rho/(2\rho)$, the only
nonzero connection coefficients are
$\Gamma^\rho{}_{\rho\rho}=-\tfrac{1}{2F}\tfrac{dF}{d\rho}$
and $\Gamma^\rho{}_{ij}=-2F\delta_{ij}$.
These extend smoothly to $\rho=0$, proving projective
compactness of order two.
The gas has energy density $\epsilon_c=3F$ and pressure
$p_c=-3/F$, approaching $3$ and $-3$, respectively, as
$\rho\to0$.
For a perfect fluid, the stress--energy tensor
is $(\epsilon+p)u_au_b+p\widetilde g_{ab}$, where~$u^a$
is the unit fluid velocity.
The constant contribution with energy density $3$ and
pressure $-3$ has stress--energy tensor
$-3\widetilde g_{ab}$.
Moving this to the left-hand side of the Einstein
equations gives the cosmological term corresponding
to $\Lambda=3$.
The remaining fluid has density $3(F-1)$ and pressure
$3(1-F^{-1})$, both vanishing at the boundary.
Its stress--energy tensor in Equation~\eqref{eq:efe}
has components
\begin{equation}\label{eq:chaplygin-stress}
 \widetilde T_{\rho\rho}=\tfrac{3b\rho}{4F(F+1)},
 \qquad
 \widetilde T_{ij}=\tfrac{3b\rho^2}{F(F+1)}\delta_{ij},
 \qquad
 \widetilde T_{\rho i}=0.
\end{equation}
Since $\tau/\rho$ is a smooth nowhere-zero density, this
gives the $q=1$ falloff of the stress--energy tensor.
\end{example}

\section{Projective Fundamental Forms}\label{sec:fundamental}

Using the boundary relations derived in
Section~\ref{sec:efe}, we construct the third and fourth
projective fundamental forms of $\Sigma$, compute their
curvature expressions, and derive the constraints imposed
by the stress--energy tensor decay.
In four dimensions, the projective fourth-form constraint
is compared with the conformal fifth-form constraint.
The final subsections examine the restrictions imposed
by energy conditions and the steps needed to construct
higher fundamental forms.

\subsection{The second fundamental form and the induced conformal geometry}
For a defining function $\rho$ of $\Sigma$, the projective second fundamental
form is represented by the restriction of $\nabla_a\nabla_b\rho$
to~$T\Sigma\times T\Sigma$. On tangent vector fields $u, v$ this equals
$-d\rho(\nabla_u v)$. A projective change adds terms with a conormal
factor, which vanish on tangential arguments. Replacing $\rho$ by
$f\rho$ with $f>0$ rescales the restricted Hessian of $\rho$ by $f|_\Sigma$. The physical metrizability solution $\lambda^{ab}$, obtained by extending
the weighted inverse $\operatorname{vol}(\widetilde g)^{2/(d+1)}
\widetilde g^{ab}$ of the physical metric, fixes the defining density
$\tau=\det\lambda$. The associated weighted second form is
\[
 \tau_{ab}:=(\nabla_a\nabla_b\tau)^\top|_\Sigma.
\]
Writing $\tau=\rho\eta$ gives
$\tau_{ab}=\eta(\nabla_a\nabla_b\rho)^\top|_\Sigma$.

Recall the weighted boundary metric $\gamma_{ab}$ introduced
in Section~\ref{sec:prelim}.
The second fundamental form satisfies
\cite[Proposition~6.3.4]{Porath2023}
\[
 \tau_{ab}\lambda^{bc}\overset{\Sigma}{=}2\alpha\Pi_a^{\top c}.
\]
Since $\lambda^{ab}$ induces the inverse boundary metric
$\gamma^{ab}$, it follows that $\tau_{ab}=2\alpha\gamma_{ab}$.
In a smooth density scale preserved by $\nabla$, the
tangential projection of $\nabla$ is the Levi--Civita
connection $\bar\nabla$ of the corresponding boundary
metric~\cite[Propositions~6.3.10--6.3.11]{Porath2023}.
In the asymptotically de Sitter setting, this metric is
positive definite.
The normalization in Section~\ref{sec:efe} gives
$\alpha|_\Sigma=-1$, and hence~$\tau_{ab}=-2\gamma_{ab}$.

\subsection{Third Fundamental Form}\label{sec:third-form}

The physical metrizability solution $\lambda^{ab}$ fixes
$\tau=\det\lambda$ and~$I_B=D_B\tau$. By Equation~\eqref{eq:dddtau},
the weight $-1$ tractor $D_CD_AI_B$ lies entirely in the $ZZZ$
slot, so the extraction $p^*(DDI)$ gives
the invariant weight 2 tensor $\Omega_{abc}$. Contracting it with the transverse vector $\nu^c$ and
restricting the remaining slots to $T\Sigma$ gives
\begin{equation}\label{eq:iii-definition}
 \III_{ab}:=\left.(\nu^c\Omega_{abc})^\top\right|_\Sigma.
\end{equation}
The weight of $\nu$ is $-2$, so $\III$ has weight zero.
With $\lambda^{ab}$ fixed, a change of representative shifts~$\nu^c$ by the tangential vector $-\lambda^{cb}\Upsilon_b$ on
$\Sigma$. Since $\Omega_{abc}$ is invariant, the change in
the pullback in Equation~\eqref{eq:iii-definition} involves only its fully
tangential part. This part vanishes under Equation~\eqref{eq:normality},
as proved in Appendix~\ref{app:third-form}, giving projective
invariance of $\III$. The calculations in Appendix~\ref{app:third-form} also give
\begin{equation}\label{eq:iii-curvature}
\III_{ab}\overset{\Sigma}{=}
 -2\alpha\Pi_a^{\top g}\Pi_b^{\top h}\nu^cW_{cg}{}^e{}_h\mu_e.
\end{equation}
We next relate $\III$ to the Schouten tensor of the
boundary metric. In a local density scale preserved
by $\nabla$, set
\[
 \underline P_{ab}:=\Pi_a^{\top c}\Pi_b^{\top e}P_{ce},
\]
and let $\bar P_{ab}$ denote the Schouten tensor of
$\gamma_{ab}$ in this scale. Then
\begin{equation}\label{eq:iii-schouten}
\III_{ab}
 =2\alpha(d-3)
 \big(
   \bar P_{ab}-\underline P_{ab}
   +\tfrac{\sigma}{2}\gamma_{ab}
 \big).
\end{equation}
Moreover, $\III_{ab}$ is trace-free with respect to
$\gamma$, so equivalently
\begin{equation}\label{eq:iii-schouten-tf}
\III_{ab}
 =2\alpha(d-3)(\bar P-\underline P)_{(ab)^{\circ}},
\end{equation}
where, for a symmetric boundary two-tensor $U$, $U_{(ab)^{\circ}}:=U_{ab} -\tfrac{1}{d-1} (\gamma^{ce}U_{ce})\gamma_{ab}$. These identities and the projective invariance of
$\III$ are proved in Appendix~\ref{app:third-form}.

The third form vanishes under the boundary normality condition in
Equation~\eqref{eq:normality}. By the prolongation identity in Equation~\eqref{eq:prolongation-middle}, the mixed slot of $\nabla_aH^{BC}$
is proportional to $W_{ac}{}^b{}_e\lambda^{ce}$. This contraction therefore vanishes on $\Sigma$, and the
commutator identity in Equation~\eqref{eq:U} reduces to
\[
 W_{ab}{}^c{}_e\lambda^{ef}
 +W_{ab}{}^f{}_e\lambda^{ce}\overset{\Sigma}{=}0.
\]
Contracting the above with $\mu_c\gamma_{fg}$ and using
$\mu_c\lambda^{ce}\overset{\Sigma}{=}0$ and
$\lambda^{ef}\gamma_{fg}\overset{\Sigma}{=}\Pi_g^{\top e}$
gives
$\mu_cW_{ab}{}^c{}_e\Pi_g^{\top e}\overset{\Sigma}{=}0$.
Substitution into Equation~\eqref{eq:iii-curvature} then yields
\begin{equation}\label{eq:iii-constraint}
 \III_{ab}\overset{\Sigma}{=}0.
\end{equation}

\subsection{Fourth Fundamental Form}\label{sec:fourth-form}
The construction of the third form uses the boundary value of~$\Omega$. The next boundary datum is its first transverse derivative. To probe this derivative in a projectively invariant way, we use the tractor Laplacian
$\Box=H^{AB}D_AD_B$, where $H^{AB}=L(\lambda)^{AB}$.
For a smooth weighted tractor field~$V$, the relation
$\lambda^{ab}\mu_b\stackrel{\Sigma}{=}0$ ensures that
$\lambda^{ab}\nabla_a\nabla_bV$ contracts the Hessian
in two tangential directions at the boundary. Thus $\Box V|_\Sigma$ involves at most one transverse
derivative of $V$. Applying~$\Box$ to the cotractor in
Equation~\eqref{eq:dddtau} gives
\[
 U_{CBA}:=\Box D_CD_AI_B.
\]
This cotractor is projectively invariant, but its
$ZZZ$ coefficient (containing the first transverse derivative of $\Omega$) depends on the splitting.
Indeed, Equation~\eqref{pre:splitting-change} shows that
$Y$ acquires a $Z$ term under the change of splitting. Hence, we first use two invariant corrections to remove the $Y$ terms from $U$.
The first, given in Equation~\eqref{eq:first-ripoff},
cancels the terms with two $Y$ factors.
The second, given in Equation~\eqref{eq:second-ripoff},
cancels the remaining terms with one $Y$ factor.
Write $\mathscr P$ for these corrections followed by
the extraction. The corrected cotractor $U^{(2)}$ has only a $ZZZ$ slot:
\begin{equation}\label{eq:H}
 U^{(2)}_{CBA}=Z_C{}^cZ_B{}^bZ_A{}^a\mathcal H_{cba},
 \qquad
 \mathcal H_{cba}:=\mathscr P(U)_{cba}.
\end{equation}
The tensor $\mathcal H_{cba}$ is projectively invariant,
symmetric in $a,b$, and has weight zero.
Contracting it with~$\nu$ and restricting to $T\Sigma$ defines the fourth fundamental form,
\begin{equation}\label{eq:iv-definition}
 \IV_{ab}:=(\nu^c\mathcal H_{cba})^\top\big|_\Sigma,
\end{equation}
which has projective weight $-2$. The fully tangential part of
$\mathcal H$ vanishes by the calculation in
Appendix~\ref{app:fourth-boundary}. Recall that for $\widehat\nabla=\nabla+\Upsilon$,
the change $\widehat\nu^c-\nu^c=-\lambda^{ca}\Upsilon_a$
is tangent to $\Sigma$. Replacing~$\nu$ by
$\widehat\nu$ changes the fourth form by
\[
\big[  (\widehat\nu^c-\nu^c)\mathcal H_{cba} \big]^{\top}\big|_{\Sigma},
\]
which vanishes because $\mathcal H^\top\overset{\Sigma}{=}0$. This proves projective invariance of $\IV$. The curvature calculation in Appendix~\ref{app:fourth-curvature} gives
\begin{equation}\label{eq:iv-final}
 \IV_{ab}\overset{\Sigma}{=}
 \tfrac{2\alpha(d-7)}{3}\mathcal E_{ab},
\end{equation}
where
\begin{equation} \label{eq:curlye}
 \mathcal E_{ab}:=
 \left.\Pi_{(a}^{\top e}\Pi_{b)}^{\top f}\nu^c
 \left(\Pi_g^{\top h}\nabla_hW_{fc}{}^g{}_e+(d-4)C_{cef}\right)
 \right|_\Sigma.
\end{equation}
We now look at the stress--energy tensor to relate $\IV$ to matter fields.
The right-hand side of Equation~\eqref{eq:omega-matter} can be promoted to a rank 3 tractor using Equation~\eqref{pre:pure-slot-extraction}. This yields
\begin{equation}\label{eq:matter-tractor}
 S_{CAB}:=Z_C{}^cZ_A{}^aZ_B{}^b
          \bigl(2\tau^{q+1}\widetilde\nabla_cS_{ab}\bigr).
\end{equation}
Here $S_{ab}$ is the unphysical trace-corrected source, and
Equation~\eqref{eq:matter-change} shows that $S_{CAB}$ extends smoothly
to $\Sigma$. Applying the tractor Laplacian and then extracting
the three-tensor gives
$\Phi_{cba}:=\mathscr P(\Box S_{CAB})_{cba}$.
Calculations in Appendix~\ref{app:matter-constraints} yield
\begin{equation}\label{eq:matter-boundary}
 \left.(\nu^c\Phi_{cba})^\top\right|_\Sigma
 =
 \begin{cases}
  4\alpha\left(\alpha(7-d)S^\top_{ab}
     +\tfrac12\tau_{ab}\lambda^{ce}S^\top_{ce}\right), & q=1,\\[3pt]
  0, & q>1 .
 \end{cases}
\end{equation}

For solutions of the Einstein equations with
$\widetilde T_{ab}=\tau^qT_{ab}$, Equations~\eqref{eq:dddtau}
and \eqref{eq:omega-matter} give $S_{CAB}=D_CD_AI_B$, so
$\Phi_{cba}=\mathcal H_{cba}$. Thus the tensor in Equation~\eqref{eq:matter-boundary} equals the fourth form $\IV_{ab}$ defined in Equation~\eqref{eq:iv-definition}.
\subsection{Summary}\label{sec:boundary-summary}

Under the stress--energy falloff $\widetilde T_{ab}=\tau^qT_{ab}$,
the Einstein equations constrain the
third and fourth fundamental forms of $\Sigma$. The following
theorem states these constraints.

\begin{theorem}\label{thm:boundary-constraints}
Let $(\widetilde M,\widetilde g)$ be an asymptotically
de Sitter spacetime of dimension $d\geq4$ admitting a
smooth projective compactification of order two with
boundary $\Sigma$.
Suppose that
\begin{equation}\label{eq:boundary-assumptions}
 \widetilde R_{ab}
 -\tfrac12\widetilde R\widetilde g_{ab}
 +\Lambda\widetilde g_{ab}
 =\widetilde T_{ab},
 \qquad
 \widetilde T_{ab}=\tau^qT_{ab},
 \qquad
 \Lambda=\tfrac{(d-1)(d-2)}2,
 \quad q\in\mathbb N_{\geq1}.
\end{equation}
Here $\tau$ is the weight~2 defining density
for $\Sigma$ determined by the physical metrizability
solution, and the unphysical stress--energy tensor
$T_{ab}$ is smooth up to $\Sigma$ and has projective
weight $-2q$.
Let $\III$ and~$\IV$ denote projective fundamental forms defined by
Equations~\eqref{eq:iii-definition} and
\eqref{eq:iv-definition}.
Then these forms are projectively invariant and satisfy
\begin{equation}\label{eq:final-constraints}
 \III_{ab}=0,
 \qquad
 \IV_{ab}=
 \begin{cases}
  -\tfrac{4(d-7)}{d-1}T^\top_{ab}, & q=1,\\[3pt]
  0, & q>1,
 \end{cases}
\end{equation}
on $\Sigma$. Here $\top$ denotes tangential restriction to $\Sigma$.
\end{theorem}
\begin{proof}
The assumed stress--energy tensor decay implies the boundary normality condition in Equation~\eqref{eq:normality}. The third form
therefore vanishes by Equation~\eqref{eq:iii-constraint}.

For the fourth form, Equations~\eqref{eq:dddtau},
\eqref{eq:omega-matter}, and \eqref{eq:matter-tractor} give
$S_{CAB}=D_CD_AI_B$. Applying $\Box$ and then $\mathscr P$
gives $\Phi_{cba}=\mathcal H_{cba}$. Hence, the left-hand side
of Equation~\eqref{eq:matter-boundary} equals the fourth form defined in Equation~\eqref{eq:iv-definition}.
For $q>1$, it follows that $\IV_{ab}=0$. For $q=1$,
Equation~\eqref{eq:matter-boundary} gives on $\Sigma$
\begin{equation} \label{eq:ivproof}
 \IV_{ab}
 =4\alpha^2(7-d)S^\top_{ab}
  +2\alpha\tau_{ab}\lambda^{ce}S^\top_{ce}.
\end{equation}
To evaluate the trace term, use Equation~\eqref{eq:matter-inverse}
to write
\[
 (d-1)S_{ab}
 =T_{ab}-\tfrac{\theta}{d-2}\zeta_{ab}
   +\tfrac{\alpha}{d-2}\tfrac{\theta}{\tau}\mu_a\mu_b.
\]
The term proportional to $\zeta_{ab}$ vanishes on $\Sigma$
by the trace condition in Equation~\eqref{eq:trace-zero}.
The quotient~$\theta/\tau$ is smooth, and the term proportional to $\mu_a\mu_b$ vanishes after tangential restriction. Thus $(d-1)S^\top_{ab}=T^\top_{ab}$ on $\Sigma$. Since
$\lambda^{ab}$ annihilates the conormal, we have
\[
 \left.\lambda^{ab}S^\top_{ab}\right|_\Sigma
 =\tfrac{1}{d-1}\left.\lambda^{ab}T^\top_{ab}\right|_\Sigma
 =\tfrac{1}{d-1}\left.\lambda^{ab}T_{ab}\right|_\Sigma
 =\tfrac{\theta|_\Sigma}{d-1}=0.
\]
The term proportional to $\tau_{ab}$ on the right-hand side of Equation~\eqref{eq:ivproof} therefore vanishes.
Substituting $S^\top_{ab}=T^\top_{ab}/(d-1)$ and
$\alpha|_\Sigma=-1$ from Section~\ref{sec:efe} into the
remaining term gives
\[
 \IV_{ab}=-\tfrac{4(d-7)}{d-1}T^\top_{ab},
\]
as claimed in Equation~\eqref{eq:final-constraints}. The invariance arguments following Equations~\eqref{eq:iii-definition}
and \eqref{eq:iv-definition} apply under the boundary identities
in Equations~\eqref{eq:normality} and \eqref{eq:omega-zero}.
Since $\Omega$, $\mathcal H$, and $\nu$ have
weights~$2$,~$0$, and $-2$, respectively, the two forms $\III$ and $\IV$ have
weights $0$ and $-2$, respectively.
\end{proof}

The factor $d-7$ in Equation~\eqref{eq:iv-final} makes the fourth form vanish
for $d=7$. Appendix~\ref{app:matter-constraints} relates~$\mathcal E_{ab}$ defined in Equation~\eqref{eq:curlye} to the leading tangential stress before
the factor $d-7$ enters. The relation holds on $\Sigma$
in every dimension $d\geq4$ and reads
\begin{equation}\label{eq:curvature-matter}
 \mathcal E_{ab}=
 \begin{cases}
  \tfrac{6}{d-1}T^\top_{ab}, & q=1,\\[3pt]
  0, & q>1.
 \end{cases}
\end{equation}

\subsection{Comparison with conformal boundary constraints}
\label{sec:conformal-comparison}

We now compare the constraints on the projective fundamental forms
with the conformal boundary constraints derived in \cite{GK2023}.
Throughout this subsection, $(\widetilde M,\widetilde g)$
is a four-dimensional asymptotically de Sitter spacetime satisfying the hypotheses of
Theorem~\ref{thm:boundary-constraints} with~$q=1$. Choose a local defining function $\rho$ for the projective
boundary, positive in the interior, such that the physical
metric takes the geodesic normal form
\[
 \widetilde g
 =-\tfrac{d\rho^2}{4\rho^2}+\tfrac{h_\rho}{\rho}.
\]
This form is available by \cite[Theorem~7]{CG2016b}
and \cite[Section~4]{FG2012}.
Here $h_\rho$ is smooth in $\rho$ up to the boundary.
We work in a smooth projective density scale in which
$\tau$ is represented by $-\rho$, so that~$h_{\rho}|_{\Sigma}=\gamma$.
For the conformal completion, take $s=\sqrt{\rho}$.
Then the conformal unphysical metric
\[
 g=s^2\widetilde g=-ds^2+h_{s^2}
\]
extends smoothly and nondegenerately to $\Sigma$. Smoothness of $h_\rho$ gives the even expansion
\[
 h_{s^2}
 =\gamma+s^2h_{(2)}+s^4h_{(4)}+O(s^6).
\]
The absence of linear and cubic terms in the above implies that the conformal
second and fourth forms vanish; see
\cite[Equations~(1.18) and~(4.38)]{GK2023}. We therefore compare the conformal third and fifth forms
with the projective third and fourth forms, respectively.

At the third order, $\III_{ab}$ is an electric contraction
of the projective Weyl tensor by
Equation~\eqref{eq:iii-curvature}, while the conformal
third form $\mathring{\mathcal K}^{(3)}_{ab}$
is proportional to the electric part of the conformal
Weyl tensor~\cite[Equation~(4.28)]{GK2023}.
In the current setting both third forms vanish:
$\III_{ab}=0$ by Equation~\eqref{eq:iii-constraint},
and~$\mathring{\mathcal K}^{(3)}_{ab}=0$ by
\cite[Equations~(4.29)--(4.30)]{GK2023}.
The two descriptions thus impose analogous electric
Weyl constraints.

To compare the projective fourth form with the conformal
fifth form, write the $q=1$ stress--energy decay assumption as
\[
 \widetilde T_{ab}=-\rho T_{ab}=-s^2T_{ab}.
\]
Thus $s^{-2}\widetilde T_{ab}$ extends smoothly to $\Sigma$,
establishing conformal decay of order two.
Its tangential boundary value is $-T^\top_{ab}$,
which is trace-free with respect to $\gamma$ by
Equation~\eqref{eq:trace-zero}. With $\Lambda=3$, Equations~(4.50) and~(4.54) of
\cite{GK2023} therefore give on~$\Sigma$,
\[
 \mathring{\mathcal K}^{(5)}_{ab}
 =6(B^\top)_{(ab)^\circ}
 =18T^\top_{ab}.
\]
Here $B_{ab}$ is the Bach tensor of $g$ and
$(B^\top)_{(ab)^\circ}$ denotes its tangential trace-free
part with respect to~$\gamma$.
Comparing this with $\IV_{ab}=4T^\top_{ab}$ from
Equation~\eqref{eq:final-constraints} yields
\begin{equation}\label{4to5}
 \IV_{ab}
 =\tfrac43(B^\top)_{(ab)^\circ}
 =\tfrac29\mathring{\mathcal K}^{(5)}_{ab}.
\end{equation}

\subsection{Null energy condition}
\label{sec:energy-conditions}
Let $(\widetilde M,\widetilde g)$ satisfy the hypotheses of
Theorem~\ref{thm:boundary-constraints}.
For $q=1$, Equation~\eqref{eq:final-constraints} relates
the fourth form to the leading tangential stress--energy tensor
$T^\top_{ab}|_\Sigma$.
We examine the restrictions on this part of $T_{ab}$
imposed by the null energy condition on the physical
stress--energy tensor~$\widetilde T_{ab}$.

The construction of the normal form of the physical metric $\widetilde{g}$ in Section~\ref{sec:conformal-comparison}
also applies in dimension~$d$.
Use the corresponding defining function $\rho$ and tangential metric $h_\rho$.
In the smooth projective density scale in which $\tau$
is represented by $-\rho$, the boundary metric is
$\gamma=h_{\rho}|_{\Sigma}$. Let $\hat n$ be the smooth vector field orthogonal to
the level sets of $\rho$ and normalized by
$d\rho(\hat n)=1$.
The geodesic normal form of the physical metric $\widetilde{g}$ gives
\[
 \widetilde g(\hat n,u)=0,
 \qquad
 \widetilde g(\hat n,\hat n)=-\tfrac{1}{4\rho^2}
\]
for every vector $u$ tangent to the level sets of $\rho$.
Since $\hat n$ is timelike, $h_\rho$ is positive definite
on these level sets.
Its nondegenerate boundary limit $\gamma$ is therefore
positive definite. Let $v$ be a $\gamma$-unit vector at a point of $\Sigma$.
Extend it smoothly to a field $\hat v$ tangent to the
level sets of $\rho$, normalized so that
$h_\rho(\hat v,\hat v)=1$.
Then $\widetilde g(\hat v,\hat v)=\rho^{-1}$, and hence
\[
 \ell=\hat v+2\sqrt{\rho}\,\hat n
\]
is null in the interior and tends to $v$ at the chosen
boundary point. In the chosen density scale, write
\[
 \widetilde T_{ab}=\rho^q t_{ab},
 \qquad
 t_{ab}:=(-1)^qT_{ab}.
\]
The tensor $t_{ab}$ is smooth up to $\Sigma$.
If $\widetilde T_{ab}$ satisfies the null energy condition
near $\Sigma$, then
\[
 \begin{aligned}
 0\leq \rho^{-q}\widetilde T_{ab}\ell^a\ell^b=t_{ab}\hat v^a\hat v^b
   +4\sqrt{\rho}\,t_{ab}\hat v^a\hat n^b+4\rho\,t_{ab}\hat n^a\hat n^b.
 \end{aligned}
\]
Taking the boundary limit gives
$t_{ab}v^av^b\geq0$.
Since $v$ was an arbitrary $\gamma$-unit vector,
$t^\top|_\Sigma$ is positive semidefinite. In the chosen density scale, $\lambda^{ab}$ induces $\gamma^{ab}$ on
$T^*\Sigma$, so Equation~\eqref{eq:trace-zero} gives
\[
 \gamma^{ab}t^\top_{ab}
 \overset{\Sigma}{=}(-1)^q\lambda^{ab}T_{ab}
 \overset{\Sigma}{=}0.
\]
All eigenvalues of $t^\top|_\Sigma$ with respect
to $\gamma$ are nonnegative, and their sum is its
$\gamma$-trace, which vanishes. Thus $t^\top|_\Sigma=0$, and consequently
\[
 T^\top_{ab}|_\Sigma=0,
 \qquad
 \IV_{ab}|_\Sigma=0,
\]
where the second equality follows from
Equation~\eqref{eq:final-constraints}.
The same conclusion holds under the weak or strong
energy condition on $\widetilde T_{ab}$ near $\Sigma$,
since each implies the null energy condition.

\subsection{Higher-order fundamental forms and constraints}
The third- and fourth-form constraints relate the projective
geometry of $\Sigma$ to the asymptotic stress--energy tensor.
Successive applications of the tractor Laplacian give a way to
extend this comparison. The transverse order of the resulting
operators determines which derivatives of the geometric data and the
matter source can enter the boundary identities. To express these
identities in terms of higher fundamental forms, one must also
extract invariant boundary two-tensors from the resulting tractor
fields. In the conformal setting, these steps lead to exceptional
dimensions and to higher forms defined subject to lower-order
vanishing conditions~\cite[Section~4.3]{BGW2023}.
Equation~\eqref{eq:differentiated-efe} takes the tractor form
$D_CD_AI_B=S_{CAB}$, where $S_{CAB}$ is given in
Equation~\eqref{eq:matter-tractor}. Applying $\Box^k$ for $k \in \mathbb{N}$ to this
identity in the interior and restricting to~$\Sigma$ gives
\begin{equation}\label{highertrac}
 \Box^kD_CD_AI_B\big|_\Sigma
 =\Box^kS_{CAB}\big|_\Sigma.
\end{equation}
The tractor-valued forms in \cite[Definition~6.5.2]{Teisseire2026}
also use successive tractor Laplacians. There the
Laplacian is built by taking two Thomas derivatives of a
normalized defining density and inverting the resulting
tensor. Here $\Box=H^{AB}D_AD_B$ is determined by the physical
metrizability solution. The coefficient in the leading transverse order term in $\Box^k$
is the same as in \cite[Proposition~5.5.3]{Teisseire2026}. In particular, for $k\geq1$,
\begin{equation}\label{leadingtransverse}
 \begin{split}
 \Box^kD_CD_AI_B\big|_\Sigma
 ={}&\bigg(\prod_{j=1}^{k}(d-4k+2j-3)\bigg) Z_C{}^cZ_A{}^aZ_B{}^b
       (\nu^e\nabla_e)^k\Omega_{abc} \big|_\Sigma\\
 &+\text{terms of lower transverse order}.
 \end{split}
\end{equation}
The remaining terms involve at most $k-1$ transverse
derivatives of $\Omega_{abc}$. To determine the coefficient in
Equation~\eqref{leadingtransverse}, take
$V=\tau^kV_k$, where $V$ has weight $-1$ and $V_k$
is smooth of weight $-1-2k$.
All transverse derivatives of $V$ of orders below $k$
vanish on $\Sigma$.
Equation~\eqref{pre:box-commutator} gives
\[
 \Box V
 =\tau^{k-1}\bigl(
   2\alpha k(d-2k-3)V_k+\tau\Box V_k
 \bigr).
\]
For each further application of $\Box$, commute it past
$\tau^i$, with $i=k-1,\ldots,1$.
The commutator term contains $\tau^{i-1}$,
while the other term retains $\tau^i$ and vanishes
on $\Sigma$ after the remaining $i-1$ applications. The final boundary value is therefore obtained
by multiplying the commutator factors
$2\alpha i(d-4k+2i-3)$ for $i=k,k-1,\ldots,1$, giving
\[
 \Box^kV\big|_\Sigma
 =
 k!(2\alpha)^k
 \bigg(\prod_{j=1}^{k}(d-4k+2j-3)\bigg)
 V_k
 \bigg|_\Sigma.
\]
Similarly, $\nu^a\nabla_a\tau=2\alpha$ on $\Sigma$
gives
\[
 \left.(\nu^e\nabla_e)^kV\right|_\Sigma
 =\left.k!(2\alpha)^kV_k\right|_\Sigma.
\]
Since $V_k|_\Sigma$ is arbitrary, comparing these
expressions and cancelling the nonzero factor
$k!(2\alpha)^k$ identifies the coefficient in
Equation~\eqref{leadingtransverse}. Whenever the product in
Equation~\eqref{leadingtransverse} is nonzero,
Equation~\eqref{highertrac} determines the $k$th transverse
derivative of $\Omega_{abc}$ from its lower transverse
derivatives and the boundary value of~$\Box^kS_{CAB}$. For $\widetilde T_{ab}=\tau^qT_{ab}$,
Equation~\eqref{eq:matter-change} gives $S_{CAB}=O(\tau^q)$.
Consequently $\Box^kS_{CAB}=O(\tau^{q-k})$ for $0\leq k\leq q$.
The right-hand side of Equation~\eqref{highertrac} therefore vanishes for $k<q$. Order $k=q$ is the first at which the
leading stress--energy data can contribute to the boundary
tractor identities.

Equation~\eqref{highertrac} gives identities between tractors.
To obtain further fundamental forms on $\Sigma$, we need to
extract a smooth, projectively invariant three-tensor. If its fully
tangential part vanishes, contraction with $\nu$ and restriction to $T\Sigma$ gives a projectively
invariant boundary two-tensor from~$\Box^k D_C D_B I_A$. With the extraction convention
used here, its weight is $-2k$. For $k=0,1$, these steps give $\III$ and $\IV$ in
Equations~\eqref{eq:iii-definition} and
\eqref{eq:iv-definition}. At higher levels, constructing
an invariant three-tensor and proving its tangential
vanishing remain open. The fourth form already uses
derivative corrections in Equations~\eqref{eq:first-ripoff}
and \eqref{eq:second-ripoff}. Such corrections can affect
the transverse order, which must therefore be checked for
the final boundary two-tensor. Dimensions in which the coefficient of the
highest transverse derivative vanishes must also be determined
separately from those of the tractor iterates. We leave these higher constructions
to future work. 

\section*{Acknowledgements}
J.K. acknowledges support from the National Science Centre, Poland, under the SONATA project No. 2025/59/D/ST2/02050. T.V. acknowledges support from the American Institute of Physics through 2025-TUTEXCEL-0005. We would also like to acknowledge the use of GPT Codex for assistance with simplifications and manuscript preparation. 

\appendix
\section{Proofs}\label{sec:proofs}

\subsection{The differentiated Einstein equation
(Section~\ref{sec:efe})}\label{app:efe-equivalence}

We prove the rigidity and equivalence statements in
Equations~\eqref{eq:parallel-rigidity} and
\eqref{eq:efe-equivalence}. Suppose that $\widetilde M$ is connected,
$\Sigma\ne\emptyset$, and $\widetilde g$ is projectively
compact of order two. Fix $p\in\Sigma$, choose a local
defining function $\rho$ positive on the interior, and use the
smooth connection $\nabla:=\widetilde\nabla+d\rho/(2\rho)$.
The metric $\widetilde g$ can be written in the following form~\cite[Theorem~7]{CG2016b},
\[
 \widetilde g_{ab}
 =\tfrac{h_{ab}}{\rho}
  +\tfrac{C}{\rho^2}\nabla_a\rho\nabla_b\rho,
 \qquad
 C=-\tfrac{d(d-1)}{4\widetilde R(p)}\ne0.
\]
Here $\widetilde R(p)$ is the boundary value of the physical
scalar curvature, $C$ is constant, and $h_{ab}$ is smooth
with nondegenerate restriction to $T\Sigma$. Substituting this metric into
$\widetilde\nabla_a\widetilde g_{bc}=0$ and using $\nabla$ gives
\[
 \begin{aligned}
 \rho\nabla_a h_{bc}
 ={}&-\bigl(C\nabla_a\nabla_b\rho+\tfrac12h_{ab}\bigr)
       \nabla_c\rho-\bigl(C\nabla_a\nabla_c\rho+\tfrac12h_{ac}\bigr)
       \nabla_b\rho.
 \end{aligned}
\]
Since $d\rho\ne0$ on $\Sigma$, the tensor
$2C\nabla_a\nabla_b\rho+h_{ab}$ vanishes there.
The transformation law of the Schouten tensor in
Equation~\eqref{eq:ptop} gives
\[
 \widetilde P_{ab}+\tfrac1{4C}\widetilde g_{ab}
 =P_{ab}
  +\tfrac{2C\nabla_a\nabla_b\rho+h_{ab}}{4C\rho}.
\]
The quotient above is smooth because its numerator vanishes
on $\Sigma$. Since the projective Weyl tensor is also
smooth, the curvature has the form
\[
 \widetilde R_{ab}{}^c{}_e
 =-\tfrac1{4C}
   \bigl(\delta_a{}^c\widetilde g_{be}
         -\delta_b{}^c\widetilde g_{ae}\bigr)
   +E_{ab}{}^c{}_e,
\]
with $E_{ab}{}^c{}_e$ smooth up to $\Sigma$.

Let $K_{ab}$ be a smooth symmetric two-tensor on
$\widetilde M$ satisfying $\widetilde\nabla_aK_{bc}=0$. Metric compatibility
gives
$\widetilde\nabla_a(\widetilde g^{bc}K_{bc})=0$,
so its trace is constant on the connected interior.
Define its trace-free part by
\[
 \mathring K_{ab}
 :=K_{ab}-\tfrac1d
   (\widetilde g^{ce}K_{ce})\widetilde g_{ab}.
\]
The constancy of the trace and metric compatibility imply
$\widetilde\nabla_a\mathring K_{bc}=0$. Our aim is to show that $K_{ab}$ is a constant multiple
of $\widetilde g_{ab}$. It therefore remains to prove that $\mathring K_{ab}=0$. Since  $\mathring K$ is parallel with respect to $\widetilde{\nabla}$, its second covariant
derivatives vanish, so
\[
 \begin{aligned}
 0
 &=[\widetilde\nabla_a,\widetilde\nabla_b]\mathring K_{ce}=-\widetilde R_{ab}{}^f{}_c\mathring K_{fe}
   -\widetilde R_{ab}{}^f{}_e\mathring K_{cf}.
 \end{aligned}
\]
Multiplying by $\widetilde g^{ac}$ yields
\[
 -\tfrac d{4C}\mathring K_{be}
 =\widetilde g^{ac}
   \bigl(E_{ab}{}^f{}_c\mathring K_{fe}
        +E_{ab}{}^f{}_e\mathring K_{cf}\bigr).
\]
In a smooth local frame near $p$, this is an algebraic
system for the components of $\mathring K$,
\[
 \left(
 -\tfrac d{4C}\delta_b{}^f\delta_e{}^g
 -\widetilde g^{ac}E_{ab}{}^f{}_c\delta_e{}^g
 -\widetilde g^{af}E_{ab}{}^g{}_e
 \right)\mathring K_{fg}=0.
\]
The coefficient matrix extends smoothly to $\Sigma$ and
equals $-\tfrac d{4C}$ times the identity at $p$, since
$\widetilde g^{-1}|_\Sigma=0$. Its determinant is therefore
nonzero at $p$ and remains nonzero nearby by continuity.
Thus $\mathring K=0$ on an interior neighbourhood of $p$.
Parallel transport on the connected interior gives
$\mathring K=0$ everywhere. Consequently,
\[
 K_{ab}
 =\tfrac1d(\widetilde g^{ce}K_{ce})\widetilde g_{ab},
\]
with constant scalar coefficient, proving
Equation~\eqref{eq:parallel-rigidity}. Applying this result to
$\widetilde P_{ab}-\widetilde S_{ab}$ proves the forward
implication in Equation~\eqref{eq:efe-equivalence}.
The reverse implication follows from metric compatibility
and the constancy of $c_0$ defined in Equation~\eqref{eq:efe-equivalence}.

\subsection{The third fundamental form
(Section~\ref{sec:third-form})}\label{app:third-form}

We first prove invariance of $\III$
defined in Equation~\eqref{eq:iii-definition}, then derive its curvature formula in Equation~\eqref{eq:iii-curvature} and intrinsic
Schouten expressions in
Equations~\eqref{eq:iii-schouten} and
\eqref{eq:iii-schouten-tf}.
Lastly, we establish the constraint in Equation~\eqref{eq:iii-constraint}. 

To prove projective invariance of $\III$, it suffices to show that
$\Omega^\top=0$. Indeed, $\Omega$ is projectively invariant,
while a change of representative shifts $\nu$ by
\[
 \widehat\nu^a-\nu^a=-\lambda^{ab}\Upsilon_b.
\]
This shift is tangential on $\Sigma$ because
$\mu_a\lambda^{ab}\overset{\Sigma}{=}0$. To show $\Omega^\top=0$, we first use
Equation~\eqref{eq:normality} to determine
$\nabla_a\mu_b$ and $\nabla_a\alpha$ on $\Sigma$.
Recall that $H_{AB}$ is the inverse of the nondegenerate
metrizability tractor $H^{AB}=L(\lambda)^{AB}$,
\[
 H_{AC}H^{CB}=\delta_A{}^B.
\]
Differentiating this relation yields $\nabla_aH_{BC} \overset{\Sigma}{=}0$. The mixed slot of that equation yields
$\nabla_a\mu_b+\beta P_{ab}-\zeta_{ab}\overset{\Sigma}{=}0$.
The identity $\beta=\tau/\alpha$ in
Equation~\eqref{pre:determinant-identities} implies
$\beta\overset{\Sigma}{=}0$, hence
\[
 \nabla_a\mu_b\overset{\Sigma}{=}\zeta_{ab}.
\]
Differentiating the determinant of $H^{BC}$ also yields
\[
 \nabla_a\alpha
 =\alpha H_{BC}\nabla_aH^{BC}
 \overset{\Sigma}{=}0.
\]
Consequently, the derivative of $\nabla_b\tau=2\alpha\mu_b$ is 
\[
 \nabla_a\nabla_b\tau
 =2(\nabla_a\alpha)\mu_b+2\alpha\nabla_a\mu_b
 \overset{\Sigma}{=}2\alpha\zeta_{ab}.
\]
Its tangential restriction is
$\tau_{ab}=2\alpha\gamma_{ab}$, so
\[
 \zeta^\top_{ab}
 \overset{\Sigma}{=}\gamma_{ab}.
\]
These identities determine the projected derivative
of the boundary projector.
Differentiating
$\Pi_f^{\top c}=\delta_f{}^c-\mu_f\nu^c$
and projecting its first two indices removes the
term containing $\mu_f$,
\begin{equation}\label{eq:projector-derivative}
 \Pi_a^{\top e}\Pi_b^{\top f}
 \nabla_e\Pi_f^{\top c}
 \overset{\Sigma}{=}-\gamma_{ab}\nu^c.
\end{equation}
Substituting $\nabla_b\tau=2\alpha\mu_b$ into
Equation~\eqref{eq:omega} and projecting all three
indices leaves
\[
 \Omega^\top_{abc}
 \overset{\Sigma}{=}
 2\alpha(\nabla_c\nabla_a\mu_b)^\top.
\]
The other terms contain $\tau$, a conormal factor,
or $\nabla\alpha$ and vanish after restriction and
projection.

To prove that the remaining second derivative vanishes,
use Equation~\eqref{pre:inverse-slots},
\[
 \lambda^{bc}\nabla_a\mu_c
 \overset{\Sigma}{=}
 \lambda^{bc}\zeta_{ac}
 =\Pi_a^{\top b}.
\]
Differentiate this boundary identity tangentially, contract with $\gamma_{bf}$ and project onto $T\Sigma$. The right-hand side vanishes by
Equation~\eqref{eq:projector-derivative} and
$\gamma_{ab}\nu^a=0$.
On the left-hand side,
Equations~\eqref{pre:metrizability} and
\eqref{pre:inverse-slots} imply
\[
 (\nabla_c\lambda^{be})\nabla_a\mu_e
 \overset{\Sigma}{=}
 -\nu^b\zeta_{ac}+\sigma\delta_c{}^b\mu_a.
\]
The same projection and contraction eliminate both
summands. In the remaining term, use~$\gamma_{ab}\lambda^{bc}
 \overset{\Sigma}{=}\Pi_a^{\top c}$ to get
\[
 (\nabla_c\nabla_a\mu_b)^\top
 \overset{\Sigma}{=}0.
\]
Thus $\Omega^\top=0$, proving projective invariance
of $\III$.

To prove Equation~\eqref{eq:iii-curvature}, expand
$\Omega_{abc}-\Omega_{cab}$ using
Equation~\eqref{eq:omega}. The commutator~$[\nabla_c,\nabla_b]\nabla_a\tau
 =R_{bc}{}^e{}_a\nabla_e\tau$ then yields
\begin{equation}\label{eq:Omega-permutation}
 \begin{aligned}
 \Omega_{abc}-\Omega_{cab}
 =R_{bc}{}^e{}_a\nabla_e\tau
 +P_{ab}\nabla_c\tau-P_{ac}\nabla_b\tau
 +2\tau C_{cba} =W_{bc}{}^e{}_a\nabla_e\tau+2\tau C_{cba}.
 \end{aligned}
\end{equation}
The second equality is obtained by canceling the Schouten parts of the curvature with the two displayed terms with $P\nabla\tau$. Contract the above with $\nu^c$ and restrict to~$T\Sigma$. The boundary identities~$\tau=0$,
$\nabla_a\tau=2\alpha\mu_a$ and $\nu^c\mu_c=1$
reduce the result to
\[
 \III_{ab}\overset{\Sigma}{=}{}
 -2\alpha\Pi_a^{\top e}\Pi_b^{\top f}
   \nu^cW_{ce}{}^g{}_f\mu_g+\bigl(
   \nu^c\nabla_a\nabla_b(2\alpha\mu_c)
   +2\alpha P_{ab}
   \bigr)^\top.
\]
For the derivative term on the right-hand side, the product rule and
$\nabla_a\alpha\overset{\Sigma}{=}0$ imply
\[
 \bigl(\nu^c\nabla_a\nabla_b(2\alpha\mu_c)\bigr)^\top
 \overset{\Sigma}{=}\bigl(\nabla_a\nabla_b(2\alpha\nu^c\mu_c)\bigr)^\top-2\alpha(\mu_c\nabla_a\nabla_b\nu^c)^\top-4\alpha(\nabla_{(a}\mu_c\nabla_{b)}\nu^c)^\top.
\]
We now simplify the right-hand side of the above display. First, Equations~\eqref{pre:inverse-slots} and
\eqref{pre:determinant-identities} imply
$
 2\alpha\nu^c\mu_c=2\alpha-2\tau\sigma.
$
The boundary identity $\nabla_b\alpha\overset{\Sigma}{=}0$
can be differentiated tangentially to yield~$
 \Pi_a^{\top c}\nabla_c\nabla_b\alpha
 \overset{\Sigma}{=}0.
$
Together with $\tau\overset{\Sigma}{=}0$ and
$(\nabla_a\tau)^\top\overset{\Sigma}{=}0$, this implies
\[
 \bigl(\nabla_a\nabla_b(2\alpha\nu^c\mu_c)\bigr)^\top
 \overset{\Sigma}{=}-2\sigma\tau_{ab}.
\]
For the second term,
Equations~\eqref{eq:prolongation-middle} and
\eqref{eq:normality} yield
\[
 W_{ac}{}^b{}_e\lambda^{ce}
 =-d\bigl(
 \nabla_a\nu^b+\delta_a{}^b\sigma-P_{ac}\lambda^{bc}
 \bigr)
 \overset{\Sigma}{=}0.
\]
The tangential derivatives of
$W_{be}{}^c{}_f\lambda^{ef}$ therefore vanish.
Differentiate Equation~\eqref{eq:prolongation-middle}
tangentially, contract with $\mu_c$, and restrict to $T\Sigma$.
The identities
$\mu_a \lambda^{ab}\overset{\Sigma}{=}0$ and
$\mu_a^\top\overset{\Sigma}{=}0$ eliminate the derivatives
of $P$ and $\sigma$, respectively. The remaining term is
\[
 \begin{aligned}
 (\mu_c\nabla_a\nabla_b\nu^c)^\top\overset{\Sigma}{=}
 (\mu_cP_{be}\nabla_a\lambda^{ce})^\top\overset{\Sigma}{=}-\underline P_{ab},
 \end{aligned}
\]
where the last equality uses
Equation~\eqref{pre:metrizability} and
$\mu_a \nu^a\overset{\Sigma}{=}1$. Finally, substitute
$\nabla_a\mu_c\overset{\Sigma}{=}\zeta_{ac}$
and Equation~\eqref{eq:prolongation-middle} into the
third term to get
\[
 (\nabla_{(a}\mu_c\nabla_{b)}\nu^c)^\top
 \overset{\Sigma}{=}
 (\zeta_{c(a}P_{b)e}\lambda^{ce})^\top
 -\sigma\gamma_{ab}\overset{\Sigma}{=}
 \underline P_{ab}-\sigma\gamma_{ab}.
\]
Here we used
$\zeta_{ab}\lambda^{bc}
 \overset{\Sigma}{=}\Pi_a^{\top c}$
from Equation~\eqref{pre:inverse-slots}. The expression for $\III_{ab}$
now reduces to the form in Equation~\eqref{eq:iii-curvature}. This form immediately implies
\[
 \gamma^{ab}\III_{ab}
 =-2\alpha\nu^c\mu_e
   W_{cf}{}^e{}_g\lambda^{fg}
 =0.
\]

To prove Equation~\eqref{eq:iii-schouten}, we compute
the Ricci tensor of the boundary metric $\gamma$ using
the Gauss equation. In a smooth density scale preserved
by $\nabla$, the projected connection $\bar\nabla$ is
the Levi--Civita connection of $\gamma$.
We first decompose ambient derivatives of tangent fields
and of $\nu$ along $\Sigma$ into tangential and transverse
components.

Let $v^a$ be tangent to $\Sigma$. Differentiating
$\mu_a v^a \overset{\Sigma}{=}0$ along $\Sigma$ gives
\[
 \mu_b\Pi_a^{\top c}\nabla_c v^b \overset{\Sigma}{=} -\Pi_a^{\top c}(\nabla_c\mu_b)v^b =-\gamma_{ab}v^b.
\]
The projection onto the transverse direction is therefore
\[
 \Pi_e^{\perp b}\Pi_a^{\top c}\nabla_c v^e =\nu^b\mu_e\Pi_a^{\top c}\nabla_c v^e \overset{\Sigma}{=}-\gamma_{ae}v^e\nu^b.
\]

For the derivative of $\nu$, substitute
$W_{ac}{}^b{}_e\lambda^{ce}\overset{\Sigma}{=}0$
into Equation~\eqref{eq:prolongation-middle}.
The transverse and tangential projections are
\[
 \Pi_a^{\top c}\Pi_e^{\perp b}\nabla_c\nu^e \overset{\Sigma}{=}0,
\]
and
\[
 \Pi_a^{\top c}\Pi_e^{\top b}\nabla_c\nu^e \overset{\Sigma}{=} \underline P_{ac}\gamma^{cb}-\sigma\Pi_a^{\top b}.
\]
Substitution into the Gauss equation yields
\[
 \bar R_{ab}{}^e{}_c =\Pi_a^{\top f}\Pi_b^{\top g}\Pi_h^{\top e}\Pi_c^{\top i}R_{fg}{}^h{}_i+\gamma_{bc}\underline P_{af}\gamma^{fe}-\gamma_{ac}\underline P_{bf}\gamma^{fe}-\sigma\bigl(\gamma_{bc}\Pi_a^{\top e}-\gamma_{ac}\Pi_b^{\top e}\bigr).
\]
The Ricci tensor is the tangential trace
\[
 \bar R_{ab}=\Pi_e^{\top c}\bar R_{ca}{}^e{}_b,
\]
 so tracing the Gauss equation gives
\[
 \bar R_{ab}\overset{\Sigma}{=}\Pi_a^{\top f}\Pi_b^{\top g}\Pi_h^{\top c}R_{cf}{}^h{}_g-\underline P_{ab}+\bigl(\gamma^{ce}\underline P_{ce}-(d-2)\sigma\bigr)\gamma_{ab}.
\]
The projective curvature decomposition and
$\Pi_a^{\top a}\overset{\Sigma}{=}d-1$ imply
\[
 \Pi_a^{\top f}\Pi_b^{\top g}\Pi_h^{\top c}R_{cf}{}^h{}_g \overset{\Sigma}{=}\Pi_a^{\top f}\Pi_b^{\top g}\Pi_h^{\top c}W_{cf}{}^h{}_g+(d-2)\underline P_{ab}.
\]
Expanding $\Pi_h^{\top c}=\delta_h{}^c-\mu_h\nu^c$ eliminates the trace of the Weyl tensor. Equation~\eqref{eq:iii-curvature}
then implies
\[
 \Pi_a^{\top f}\Pi_b^{\top g}\Pi_h^{\top c}W_{cf}{}^h{}_g \overset{\Sigma}{=}-\Pi_a^{\top f}\Pi_b^{\top g}\nu^c\mu_hW_{cf}{}^h{}_g=\tfrac1{2\alpha}\III_{ab}.
\]
Combining these expressions gives
\[
 \bar R_{ab}=\tfrac1{2\alpha}\III_{ab}+(d-3)\underline P_{ab}+\bigl(\gamma^{ce}\underline P_{ce}-(d-2)\sigma\bigr)\gamma_{ab}.
\]
Taking the $\gamma$-trace and using
$\gamma^{ab}\III_{ab}=0$, we obtain
\[
 \tfrac{\bar R}{2(d-2)}=\gamma^{ce}\underline P_{ce}-\tfrac{d-1}{2}\sigma.
\]
The Schouten tensor of the $(d-1)$-dimensional boundary
therefore satisfies
\[
 \bar P_{ab}=\tfrac1{d-3}\left(\bar R_{ab}-\tfrac{\bar R}{2(d-2)}\gamma_{ab}\right)=\underline P_{ab}-\tfrac{\sigma}{2}\gamma_{ab}+\tfrac1{2\alpha(d-3)}\III_{ab}.
\]
Solving for $\III_{ab}$ proves
Equation~\eqref{eq:iii-schouten}.
As $\III$ is trace-free, taking the trace-free part of
Equation~\eqref{eq:iii-schouten} proves
Equation~\eqref{eq:iii-schouten-tf}.

To prove Equation~\eqref{eq:iii-constraint}, it suffices,
by Equation~\eqref{eq:iii-curvature}, to establish
\[
 \mu_eW_{ca}{}^e{}_g\Pi_b^{\top g}\overset{\Sigma}{=}0.
\]
Boundary normality implies that
$W_{ac}{}^b{}_e\lambda^{ce}\stackrel{\Sigma}{=}0$.
To connect these two parts of $W$ on $\Sigma$, we use the
metrizability Equation~\eqref{pre:metrizability}. Differentiating it and commuting derivatives give
\[
 [\nabla_a,\nabla_b]\lambda^{ef}=-\delta_b{}^e\nabla_a\nu^f-\delta_b{}^f\nabla_a\nu^e+\delta_a{}^e\nabla_b\nu^f+\delta_a{}^f\nabla_b\nu^e.
\]
The curvature formula expresses the same quantity as
\[
 [\nabla_a,\nabla_b]\lambda^{ef}=R_{ab}{}^e{}_g\lambda^{gf}+R_{ab}{}^f{}_g\lambda^{eg}.
\]
Replace the derivatives of $\nu$ in the first expression
using Equation~\eqref{eq:prolongation-middle}, and
decompose $R$ into its Weyl and Schouten parts in the
second. The $\sigma$ terms cancel in the first
expression, while the Schouten terms agree in both.
The remaining terms therefore satisfy
\begin{equation}\label{eq:U}
 W_{ab}{}^e{}_g\lambda^{gf}+W_{ab}{}^f{}_g\lambda^{eg}=\tfrac1d\lambda^{cg}\bigl(\delta_b{}^eW_{ac}{}^f{}_g+\delta_b{}^fW_{ac}{}^e{}_g-\delta_a{}^eW_{bc}{}^f{}_g-\delta_a{}^fW_{bc}{}^e{}_g\bigr).
\end{equation}
Every term on the right-hand side vanishes
on $\Sigma$ by Equation~\eqref{eq:normality}. Contracting the above with~$\mu_e\gamma_{fh}$ and using $\mu_a\lambda^{ab}\overset{\Sigma}{=}0$ and
$\gamma_{ab}\lambda^{ac}\overset{\Sigma}{=}\Pi_b^{\top c}$ therefore yields
\[
 0\overset{\Sigma}{=}\mu_e\gamma_{fb}\bigl(W_{ca}{}^e{}_g\lambda^{gf}+W_{ca}{}^f{}_g\lambda^{eg}\bigr)=\mu_eW_{ca}{}^e{}_g\Pi_b^{\top g}.
\]
Equation~\eqref{eq:iii-curvature} now implies
\[
 \III_{ab}\overset{\Sigma}{=}-2\alpha\Pi_a^{\top f}\nu^c\mu_eW_{cf}{}^e{}_g\Pi_b^{\top g}=0,
\]
proving Equation~\eqref{eq:iii-constraint}.
\subsection{Tractor construction of the fourth fundamental form (Section~\ref{sec:fourth-form})}\label{app:fourth-slots}

We construct the corrected cotractor in
Equation~\eqref{eq:H} by removing all terms containing
$Y$ from $U_{CBA}:=\Box D_CD_AI_B$.
This leaves a pure $ZZZ$ cotractor, whose coefficient
is the invariant tensor $\mathcal H_{cba}$.
The construction has two stages: removing the terms
with two $Y$ factors, then those with one.

We first compute the slots of $U$.
The cotractor in Equation~\eqref{eq:dddtau} has weight
$-1$, so its first Thomas derivative acts as
$D_D=-Y_D+Z_D{}^e\nabla_e$. Applying a second Thomas derivative at weight~$-2$, using~$\nabla_a Z_A{}^b=-\delta_a{}^bY_A$ and
$\nabla_a Y_A=P_{ab}Z_A{}^b$, and contracting with $H^{DE}$
produces
\begin{equation}\label{eq:raw-box-ddd}
\begin{aligned}
 U_{CBA}= & B_aY_CY_BZ_A{}^a+B_bY_CZ_B{}^bY_A+D_cZ_C{}^cY_BY_A+E_{ba}Y_CZ_B{}^bZ_A{}^a \\
  & +F_{ca}Z_C{}^cY_BZ_A{}^a+F_{cb}Z_C{}^cZ_B{}^bY_A+H_{cba}Z_C{}^cZ_B{}^bZ_A{}^a.
 \end{aligned}
\end{equation}
The coefficients are
\[
 B_a:=2\lambda^{ef}\Omega_{aef},\qquad D_c:=2\lambda^{ef}\Omega_{efc},
\]
\[
 E_{ba}:=4\nu^e\Omega_{abe}-2\lambda^{ef}\nabla_e\Omega_{abf},
\]
\[
 F_{ca}:=4\nu^e\Omega_{aec}-2\lambda^{ef}\nabla_e\Omega_{afc},
\]
and
\[
 H_{cba}:=\lambda^{ef}\bigl(\nabla_f\nabla_e\Omega_{abc}-P_{fe}\Omega_{abc}-P_{fc}\Omega_{abe}-P_{fb}\Omega_{aec}-P_{fa}\Omega_{ebc}\bigr)-4\nu^e\nabla_e\Omega_{abc}+2\sigma\Omega_{abc}.
\]
We next remove the three terms containing two $Y$
factors. Their coefficients are invariant: under the
splitting change in Equation~\eqref{pre:splitting-change},
only a $YYY$ term could change these coefficients,
and that term is absent.
Define
\[
 \xi_a:=\tfrac13(2B_a+D_a),\qquad \chi_a:=\tfrac12(B_a-D_a),
\]
so that
\[
 B_a=\xi_a+\tfrac23\chi_a,\qquad D_a=\xi_a-\tfrac43\chi_a.
\]
We use $\xi$ and $\chi$ to construct the invariant
auxiliary cotractors
\[
 S_A:=\xi_aZ_A{}^a,\qquad Q_{CA}:=X^BU_{[C|B|A]}.
\]
The invariance of $S_A$ follows from that of $\xi$ and
$Z$; the invariance of $Q_{CA}$ follows directly from
its definition. Their weights are $-1$ and $-2$,
respectively. The splitting of $Q$ is
\[
 Q_{CA}=\chi_aY_CZ_A{}^a-\chi_cZ_C{}^cY_A+F_{[ca]}Z_C{}^cZ_A{}^a.
\]
The first differential correction of $U$ is
\begin{equation}\label{eq:first-ripoff}
 U^{(1)}_{CBA}:=U_{CBA}-\tfrac12D_CD_{(B}S_{A)}+\tfrac13(D_BQ_{CA}+D_AQ_{CB}).
\end{equation}
The choices of $\xi$ and $\chi$ cancel all three terms
containing two $Y$ factors in $U^{(1)}$. No $YYY$ term is introduced, as two Thomas derivatives
of the pure $Z$ cotractor $S_A$ produce at most two
$Y$ factors, as does one derivative of $Q_{CA}$. The one-$Y$ and the $ZZZ$ coefficients in $U^{(1)}$ are
\[
 E^{(1)}_{ab}=E_{ab}+\nabla_{(a}\xi_{b)}+\tfrac23\nabla_{(a}\chi_{b)},
\]
\[
 F^{(1)}_{ca}=F_{ca}+\tfrac34\nabla_c\xi_a+\tfrac14\nabla_a\xi_c-F_{[ca]}-\tfrac13\nabla_a\chi_c,
\]
and
\[
 H^{(1)}_{cba}=H_{cba}-\tfrac12\nabla_c\nabla_{(b}\xi_{a)}+\xi_{(a}P_{b)c}+\tfrac13\bigl(\nabla_bF_{[ca]}+\nabla_aF_{[cb]}+\chi_aP_{bc}+\chi_bP_{ac}-2\chi_cP_{ab}\bigr).
\]
With the higher slots removed, the one-$Y$ coefficients
are invariant. We now remove these last $Y$ terms.
Introduce pure $ZZ$ cotractors
\[
 K_{AB}:=p_{ab}Z_A{}^a Z_B{}^b,\qquad L_{AB}:=q_{ab}Z_A{}^a Z_B{}^b,
\]
where $p_{ab}=p_{(ab)}$ and $q_{ab}$ are invariant tensors, and set
\begin{equation}\label{eq:second-ripoff}
 U^{(2)}_{CBA}:=U^{(1)}_{CBA}-D_CK_{BA}-D_BL_{CA}-D_AL_{CB}.
\end{equation}
The two independent one-$Y$ coefficients vanish precisely
when
\[
 E^{(1)}_{ab}+2p_{ab}+2q_{(ab)}=0,\qquad F^{(1)}_{ab}+p_{ab}+3q_{ab}=0.
\]
Solving for $p$ and $q$ gives
\[
 p_{ab}=-\tfrac34E^{(1)}_{ab}+\tfrac12F^{(1)}_{(ab)},\qquad q_{ab}=\tfrac14E^{(1)}_{ab}-\tfrac12F^{(1)}_{(ab)}-\tfrac13F^{(1)}_{[ab]}.
\]
Thomas derivative of a pure $ZZ$ cotractor introduces
at most one $Y$ factor, so the previously removed slots with more than one $Y$ remain zero. The corrected cotractor is therefore
\[
 U^{(2)}_{CBA}=Z_C{}^cZ_B{}^bZ_A{}^a\mathcal H_{cba},
\]
where
\[
 \mathcal H_{cba}=H^{(1)}_{cba}-\nabla_cp_{ba}-\nabla_bq_{ca}-\nabla_aq_{cb}.
\]
The invariant inclusion in
Equation~\eqref{pre:pure-slot-extraction} identifies
$\mathcal H$ as a projectively invariant tensor.
It has weight zero and is symmetric in $a,b$.

For the subsequent boundary calculations, we expand
$\mathcal H$ by substituting the preceding coefficients
and replacing derivatives of $\lambda$ and $\nu$ using
Equations~\eqref{pre:metrizability} and
\eqref{eq:prolongation-middle}.
The result is
\[
 \mathcal H_{cba}=\widetilde{\mathcal D}_{cba}+\widetilde{\mathcal V}_{cba}+\widetilde{\mathcal W}_{cba}+\widetilde{\mathcal S}_{cba}.
\]
The four terms collect second derivatives of $\Omega$,
first derivatives multiplied by $\nu$, Weyl curvature
terms, and terms involving $P$ or $\sigma$, respectively:

\begin{equation}\label{eq:H-D}
\begin{split}
 \widetilde{\mathcal D}_{cba}=\lambda^{ef}\Bigl(&\nabla_f\nabla_e\Omega_{abc}
 -\tfrac32\nabla_c\nabla_e\Omega_{abf}  +\nabla_c\nabla_e\Omega_{f(ab)}+\nabla_{(a}\nabla_{|e|}\Omega_{b)cf} \\
& -\tfrac53\nabla_{(a}\nabla_{|e|}\Omega_{b)fc}
 -\tfrac13\nabla_{(a}\nabla_{|e|}\Omega_{|cf|b)}+\tfrac13\nabla_c\nabla_{(a}\Omega_{b)ef}  -\tfrac56\nabla_c\nabla_{(a}\Omega_{|ef|b)}\\
& +\tfrac12\nabla_b\nabla_{(c}\Omega_{|ef|a)}
 +\tfrac12\nabla_a\nabla_{(c}\Omega_{|ef|b)}+\tfrac13\nabla_b\nabla_{[c}\Omega_{a]ef}
 +\tfrac13\nabla_a\nabla_{[c}\Omega_{b]ef}
 \Bigr).
\end{split}
\end{equation}

\begin{equation}\label{eq:H-V}
\begin{split}
 \widetilde{\mathcal V}_{cba}=\nu^e\Bigl(\tfrac{23}{6}\nabla_c\Omega_{abe}
 -3\nabla_c\Omega_{e(ab)}
 -3\nabla_{(a}\Omega_{b)ce}+\tfrac73\nabla_{(a}\Omega_{b)ec}
 +\tfrac73\nabla_{(a}\Omega_{|ce|b)}
 -\tfrac52\nabla_e\Omega_{(abc)}
 \Bigr).
\end{split}
\end{equation}

\begin{equation}\label{eq:H-W}
\begin{split}
 \widetilde{\mathcal W}_{cba}={}&
 -\tfrac1dW_{ce}{}^g{}_f\lambda^{ef}
 \bigl(\tfrac83\Omega_{abg}-\tfrac23\Omega_{g(ab)}\bigr)-\tfrac1d\lambda^{ef}W_{ae}{}^g{}_f
 (\Omega_{bgc}-\Omega_{bcg})\\
 & -\tfrac1d\lambda^{ef}W_{be}{}^g{}_f
 (\Omega_{agc}-\Omega_{acg}).
\end{split}
\end{equation}

\begin{equation}\label{eq:H-S}
\begin{split}
 \widetilde{\mathcal S}_{cba}={}&
 -P_{ef}\lambda^{ef}\Omega_{abc}
 +P_{cf}\lambda^{ef}
 \bigl(\tfrac53\Omega_{abe}-\tfrac23\Omega_{e(ab)}\bigr)-\lambda^{ef}
 \bigl(P_{af}\Omega_{bce}+P_{bf}\Omega_{ace}\bigr) \\
  &+\lambda^{ef}
 \bigl(\Omega_{aef}P_{bc}+\Omega_{bef}P_{ac}\bigr)-\tfrac23\lambda^{ef}
 (\Omega_{cef}-\Omega_{efc})P_{ab}+\tfrac83\sigma
 (\Omega_{c(ab)}-\Omega_{abc}).
\end{split}
\end{equation}

\subsection{Boundary identities for the fourth fundamental form (Section~\ref{sec:fourth-form})}\label{app:fourth-boundary}

To prove projective invariance of $\IV$, it suffices
to show $\mathcal H^\top\overset{\Sigma}{=}0$:
$\mathcal H$ is invariant by construction, and a change
of representative shifts $\nu$ tangentially in
Equation~\eqref{eq:iv-definition}.
We use the boundary identities established in
Equations~\eqref{eq:normality} and
\eqref{eq:omega-zero}. 

All the terms in Equations~\eqref{eq:H-W} and
\eqref{eq:H-S} vanish with $\Omega$.
It remains to consider the first and second derivatives
of $\Omega$ in Equations~\eqref{eq:H-V} and
\eqref{eq:H-D}.
We first express their tangential projections in terms
of $Q^\top$, where $Q_{abc}:=\nu^e\nabla_e\Omega_{abc}$, then prove that $Q^\top$ vanishes. The tangential derivatives of $\Omega$ vanish on $\Sigma$,
so
\[
 \nabla_e\Omega_{abc}\overset{\Sigma}{=}\mu_eQ_{abc}.
\]
Differentiating
$\Pi_f^{\top e}\nabla_e\Omega_{abc}\overset{\Sigma}{=}0$
along $\Sigma$, and using Equation~\eqref{eq:projector-derivative} yields
\begin{equation}\label{eq:master-tt}
 \Pi_e^{\top g}\Pi_f^{\top h}\nabla_g\nabla_h\Omega_{abc}\overset{\Sigma}{=}\gamma_{ef}Q_{abc}.
\end{equation}

These identities explain the role of $Q^\top$ in  $\mathcal H^\top$.
In the tangential projection of
Equation~\eqref{eq:H-V}, tangential derivatives of
$\Omega$ vanish, while differentiation along $\nu$
leaves $Q^\top$.
In Equation~\eqref{eq:H-D}, both derivatives are tangential after projection and contraction with
$\lambda$, because
$\lambda^{ab}\mu_b\overset{\Sigma}{=}0$.
Equation~\eqref{eq:master-tt} then reduces every term
to a contraction of $Q^\top$.
Thus $Q^\top=0$ will imply $\mathcal H^\top=0$.

We have
\[
 Q^\top_{abc}\overset{\Sigma}{=}\nu^e\Pi_a^{\top f}\Pi_b^{\top g}\Pi_c^{\top h}\bigl(\nabla_e\Omega_{fgh}-\nabla_h\Omega_{fge}\bigr).
\]
Expand using Equation~\eqref{eq:omega} and decompose
the curvature commutator.
The $\tau$ and conormal terms vanish on $\Sigma$,
while the $P$ and $\nabla P$ terms cancel, leaving
\[
 Q^\top_{abc}\overset{\Sigma}{=}-\bigl(\nu^eW_{ec}{}^f{}_a\tau_{fb}+\nu^eW_{ec}{}^f{}_b\tau_{af}\bigr)^\top.
\]
The Hessian factors reduce to $\tau_{ab}$ because
$\nabla_a\nabla_b\tau\overset{\Sigma}{=}2\alpha\zeta_{ab}$
and $\nu^a\zeta_{ab}=-\sigma\mu_b$. To show that the right-hand side of the above display vanishes, rewrite it using $\tau_{ab}=2\alpha\gamma_{ab}$ and
$\gamma_{ab}\lambda^{bc}\overset{\Sigma}{=}\Pi_a^{\top c}$. This gives
\[
 Q^\top_{abc}\overset{\Sigma}{=}-2\alpha\Pi_c^{\top i}\nu^e\gamma_{af}\gamma_{bg}\bigl(W_{ei}{}^f{}_h\lambda^{hg}+W_{ei}{}^g{}_h\lambda^{fh}\bigr).
\]
The sum in parentheses is the left-hand side of
Equation~\eqref{eq:U}. Its right-hand side vanishes
on $\Sigma$ by Equation~\eqref{eq:normality}.
Therefore $Q^\top\overset{\Sigma}{=}0$, so $\mathcal H^\top\overset{\Sigma}{=}0$ and  hence $\IV$ is projectively invariant. 

We now compute $\IV$. Since $\Omega$ vanishes on $\Sigma$,
the terms $\widetilde{\mathcal W}$ and
$\widetilde{\mathcal S}$ in Equations~\eqref{eq:H-W} and \eqref{eq:H-S}
vanish there. The remaining two terms give
\[
 \IV_{ab}\overset{\Sigma}{=}(\nu^c\widetilde{\mathcal V}_{cba})^\top+(\nu^c\widetilde{\mathcal D}_{cba})^\top.
\]
For the terms involving first derivatives of $\Omega$,
substitute
$\nabla_e\Omega_{abc}\overset{\Sigma}{=}\mu_eQ_{abc}$
into Equation~\eqref{eq:H-V}.
After contraction with $\nu^c$ and tangential projection,
only the first, second and last terms do not vanish. The symmetry of $Q$ implies
\[
 (\nu^cQ_{(abc)})^\top=\tfrac13\bigl(\nu^c(Q_{abc}+2Q_{c(ab)})\bigr)^\top,
\]
and therefore
\begin{equation}\label{eq:V-boundary}
 (\nu^c\widetilde{\mathcal V}_{cba})^\top\overset{\Sigma}{=}\bigl(\nu^c(3Q_{abc}-\tfrac{14}{3}Q_{c(ab)})\bigr)^\top.
\end{equation}

We next evaluate the second derivatives of $\Omega$ in
Equation~\eqref{eq:H-D}. After contraction with $\nu^c$ and tangential projection,
each term contains either two tangential derivatives
or one tangential and one transverse derivative. Equation~\eqref{eq:master-tt} handles the first case.
For the mixed derivatives, the definition of $Q$ gives
\[
 \Pi_e^{\top f}\nabla_fQ_{abc}\overset{\Sigma}{=}\Pi_e^{\top f}\nu^g\nabla_f\nabla_g\Omega_{abc}\overset{\Sigma}{=}\Pi_e^{\top f}\nu^g\nabla_g\nabla_f\Omega_{abc}.
\]
The product-rule term
$\Pi_e^{\top f}(\nabla_f\nu^g)\nabla_g\Omega_{abc}$
vanishes because differentiation of $\nu$ along $\Sigma$
is tangential, and tangential derivatives of $\Omega$
are zero.
The interchange of derivatives adds curvature acting
on $\Omega$, which also vanishes on $\Sigma$. The mixed terms thus require the projected derivative of $Q$.
To evaluate this derivative, differentiate
the boundary identity
\[
 \Pi_a^{\top e}\Pi_b^{\top f}\Pi_c^{\top g}Q_{efg}\overset{\Sigma}{=}0.
\]
Tangential differentiation acts on $Q$ and on each
of the three projectors.
Using Equation~\eqref{eq:projector-derivative} and
moving the three projector terms to the right gives
\begin{equation}\label{eq:master-Q}
 \Pi_e^{\top f}\Pi_a^{\top g}\Pi_b^{\top h}\Pi_c^{\top i}\nabla_fQ_{ghi}\overset{\Sigma}{=}\gamma_{ea}(\nu^fQ_{fbc})^\top+\gamma_{eb}(\nu^fQ_{fac})^\top+\gamma_{ec}(\nu^fQ_{abf})^\top.
\end{equation}
We now apply these identities to the terms in
Equation~\eqref{eq:H-D}. This yields
\begin{equation}\label{eq:D-boundary}
 (\nu^c\widetilde{\mathcal D}_{cba})^\top\overset{\Sigma}{=}\bigl(\nu^c(-\tfrac{d-1}{2}Q_{abc}+(d-\tfrac73)Q_{c(ab)})\bigr)^\top+\tfrac{\gamma_{ab}}6\nu^c\lambda^{ef}(Q_{efc}+2Q_{cef}).
\end{equation}
Adding the contribution from $\widetilde{\mathcal V}$
in Equation~\eqref{eq:V-boundary} gives the following
expression for $\IV$:
\begin{equation}\label{eq:IV-AB}
 \IV_{ab}\overset{\Sigma}{=}\tfrac{7-d}{2}\bigl(\nu^c(Q_{abc}-2Q_{c(ab)})\bigr)^\top+\tfrac{\gamma_{ab}}6\nu^c\lambda^{ef}(Q_{efc}+2Q_{cef}).
\end{equation}

\subsection{Curvature formula for the fourth fundamental form
(Section~\ref{sec:fourth-form})}\label{app:fourth-curvature}

We derive Equation~\eqref{eq:iv-final} by expressing the
contractions of $Q$ in Equation~\eqref{eq:IV-AB} in terms
of the curvature tensor~$\mathcal E_{ab}$ defined in
Equation~\eqref{eq:curlye}.
The combination of $Q$ in the first term of
Equation~\eqref{eq:IV-AB} can be written~as
\begin{equation}\label{eq:Sigma-K}
 \bigl(\nu^c(Q_{abc}-2Q_{c(ab)})\bigr)^\top=\tfrac43\bigl(\nu^c(Q_{abc}-Q_{c(ab)})\bigr)^\top-(\nu^cQ_{(abc)})^\top.
\end{equation}

To simplify calculations, we first record the consequences of
Equation~\eqref{eq:normality}:
\[
 W_{ab}{}^e{}_c\lambda^{bc}\overset{\Sigma}{=}0,\qquad C_{abc}\lambda^{bc}\overset{\Sigma}{=}0.
\]
Since these tensors vanish along $\Sigma$, their
covariant derivatives in tangential directions vanish
there as well. Equation~\eqref{eq:Omega-permutation}, with
$\Omega|_\Sigma=0$, $\tau|_\Sigma=0$ and
$\nabla_e\tau=2\alpha\mu_e$, gives
\[
 W_{ab}{}^e{}_c\mu_e\overset{\Sigma}{=}0.
\]
To evaluate the first contraction on the right-hand
side of Equation~\eqref{eq:Sigma-K}, differentiate
Equation~\eqref{eq:Omega-permutation} along $\nu$,
contract with $\nu^c$, and take the symmetric
tangential part. Using $\nu^a \nabla_a \tau=2\alpha$ gives
\[
 \bigl(\nu^c(Q_{abc}-Q_{c(ab)})\bigr)^\top\overset{\Sigma}{=}2\alpha\bigl(\nu^c\mu_e\nu^f\nabla_fW_{(b|c|}{}^e{}_{a)}+2\nu^cC_{c(ab)}\bigr)^\top.
\]
The product-rule term containing the derivative of
$\nabla \tau$ vanishes because
$\nu^a\nabla_a\nabla_b\tau\overset{\Sigma}{=}-2\alpha\sigma\mu_b$
and~$W_{ab}{}^e{}_c\mu_e\overset{\Sigma}{=}0$.
The remaining normal derivative of $W$ is expressed
through its tangential divergence and the Cotton tensor
using Equation~\eqref{pre:contracted-bianchi}.
Indeed, the decomposition
$\delta_a{}^b=\Pi_a^{\top b}+\mu_a\nu^b$ gives
\[
 \bigl(\nu^c\Pi_e^{\top f}\nabla_fW_{(b|c|}{}^e{}_{a)}+\nu^c\mu_e\nu^f\nabla_fW_{(b|c|}{}^e{}_{a)}\bigr)^\top\overset{\Sigma}{=}(2-d)(\nu^cC_{c(ab)})^\top.
\]
Substituting this relation into the preceding expression
produces the combination defining $\mathcal E_{ab}$ in
Equation~\eqref{eq:curlye}. Thus
\begin{equation}\label{eq:K-curvature}
 \bigl(\nu^c(Q_{abc}-Q_{c(ab)})\bigr)^\top\overset{\Sigma}{=}-2\alpha\mathcal E_{ab}.
\end{equation}

For the last term on the right-hand side in
Equation~\eqref{eq:Sigma-K}, we differentiate the fully
symmetrized formula for $\Omega$. Equation~\eqref{eq:omega}
gives
\[
 \Omega_{(abc)}=\nabla_{(a}\nabla_b\nabla_{c)}\tau+4P_{(ab}\nabla_{c)}\tau+2\tau\nabla_{(a}P_{bc)}.
\]
Applying $\nu^f\nabla_f$ to the above, contracting with $\nu^c$ and
projecting onto directions tangential to $\Sigma$ gives~$(\nu^cQ_{(abc)})^\top$ on the left-hand side. The contribution to $(\nu^cQ_{(abc)})^\top$ containing
four derivatives of $\tau$ is
\[
 \bigl(\nu^c\nu^f\nabla_f\nabla_{(a}\nabla_b\nabla_{c)}\tau\bigr)^\top.
\]
To evaluate it, we rewrite the derivatives so that
the tangential derivative selected by $\Pi_a^{\top e}$
acts on three derivatives of $\tau$. This allows us
to differentiate the boundary formula for those derivatives along $\Sigma$.
First expand the symmetrization, then commute
$\nabla_f$ and $\nabla_a$:
\[
 \nabla_{(a}\nabla_b\nabla_{c)}\tau=\nabla_a\nabla_b\nabla_c\tau+\tfrac13(R_{ab}{}^e{}_c+R_{ac}{}^e{}_b)\nabla_e\tau,
\]
\[
 \nabla_f\nabla_a\nabla_b\nabla_c\tau=\nabla_a\nabla_f\nabla_b\nabla_c\tau-R_{fa}{}^e{}_b\nabla_e\nabla_c\tau-R_{fa}{}^e{}_c\nabla_b\nabla_e\tau.
\]
After tangential projection, the first term on the
right has the required form
$\Pi_a^{\top e}\nabla_e(\nabla_f\nabla_b\nabla_c\tau)$. We now simplify this term. Equation~\eqref{eq:omega}
and $\Omega|_\Sigma=0$ give
\[
 \nabla_f\nabla_b\nabla_c\tau\overset{\Sigma}{=}-2\alpha(2\mu_fP_{bc}+\mu_bP_{fc}+\mu_cP_{fb}).
\]
Differentiating tangentially and using
$\nabla_e\alpha\overset{\Sigma}{=}0$ yields
\[
 \Pi_a^{\top e}\nabla_e\nabla_f\nabla_b\nabla_c\tau\overset{\Sigma}{=}-2\alpha\Pi_a^{\top e}\nabla_e(2\mu_fP_{bc}+\mu_bP_{fc}+\mu_cP_{fb}).
\]
Lastly, we evaluate the derivatives of $\mu$ and the Hessians
of $\tau$ using
$\nabla_a\mu_b\overset{\Sigma}{=}\zeta_{ab}$ and
$\nabla_a\nabla_b\tau\overset{\Sigma}{=}2\alpha\zeta_{ab}$. This expresses the contribution containing four
derivatives of $\tau$ through boundary tensors,
curvature and its first derivatives.

We now add the contributions from the three summands of $\Omega_{(abc)}$. Terms with derivatives of $P$ combine to yield
\[
 \tfrac{10\alpha}{3}\bigl(\nu^c(\nabla_cP_{ab}-\nabla_{(a}P_{b)c})\bigr)^\top=\tfrac{10\alpha}{3}(\nu^cC_{c(ab)})^\top.
\]
The contributions proportional to $\sigma\underline P_{ab}$
and $\gamma_{ab}\nu^c\nu^eP_{ce}$ cancel.
Equation~\eqref{pre:contracted-bianchi} then replaces
the remaining normal derivative of $W$ by its
tangential divergence and the Cotton tensor.
Using the definition of $\mathcal E_{ab}$ in
Equation~\eqref{eq:curlye} gives
\begin{equation}\label{eq:Sigma-curvature}
 (\nu^cQ_{(abc)})^\top\overset{\Sigma}{=}2\alpha\bigl(\nu^cC_{c(ab)}-\nu^c\nu^eW_{c(a}{}^f{}_e\gamma_{b)f}\bigr)^\top-\tfrac{2\alpha}{3}\mathcal E_{ab}.
\end{equation}
Together with Equation~\eqref{eq:K-curvature},
this expresses all terms in Equation~\eqref{eq:Sigma-K} through curvature.

Before substituting them into the formula for $\IV$,
we evaluate the trace term in
Equation~\eqref{eq:IV-AB}.
Equation~\eqref{eq:Sigma-curvature} reduces this computation to
the traces of three curvature terms on its right-hand side.
The Cotton term has zero trace by
$C_{abc}\lambda^{bc}\overset{\Sigma}{=}0$.
For $\mathcal E$, tangential differentiation of the
vanishing Weyl contraction gives
\[
 0\overset{\Sigma}{=}\nu^a\Pi_b^{\top c}\nabla_c(W_{ae}{}^b{}_f\lambda^{ef}).
\]
Expanding the derivative and using the definition in
Equation~\eqref{eq:curlye} therefore yields
\[
 \lambda^{ab}\mathcal E_{ab}\overset{\Sigma}{=}\nu^a\Pi_b^{\top c}W_{ae}{}^b{}_f\nabla_c\lambda^{ef}=0.
\]
For the last equality, use
Equation~\eqref{pre:metrizability}. The undifferentiated Weyl term also has zero trace:
\[
 \lambda^{ab}\bigl(\nu^c\nu^eW_{c(a}{}^f{}_e\gamma_{b)f}\bigr)^\top\overset{\Sigma}{=}\nu^c\nu^eW_{ca}{}^a{}_e-\nu^c\nu^e\nu^a\mu_fW_{ca}{}^f{}_e=0.
\]
Taking the trace of
Equation~\eqref{eq:Sigma-curvature} consequently gives
\[
 \nu^c\lambda^{ef}(Q_{efc}+2Q_{cef})=3\nu^c\lambda^{ef}Q_{(efc)}\overset{\Sigma}{=}0.
\]
Thus the trace term in Equation~\eqref{eq:IV-AB}
vanishes. Combining Equations~\eqref{eq:Sigma-K},
\eqref{eq:K-curvature} and \eqref{eq:Sigma-curvature}
therefore gives
\[
 \IV_{ab}\overset{\Sigma}{=}\alpha(7-d)\bigl((\nu^c\nu^eW_{c(a}{}^f{}_e\gamma_{b)f}-\nu^cC_{c(ab)})^\top-\mathcal E_{ab}\bigr).
\]
The last step is to express $\IV$ entirely through
$\mathcal E$. By Equation~\eqref{eq:normality}, the tensor on the
left-hand side of Equation~\eqref{eq:U} vanishes
along $\Sigma$. Its tangential divergence therefore gives
\[
 0\overset{\Sigma}{=}\Pi_e^{\top f}\nabla_f(W_{ab}{}^c{}_g\lambda^{ge}+W_{ab}{}^e{}_g\lambda^{cg}).
\]
To obtain a symmetric boundary two-tensor, contract
with $\nu^a\gamma_{ch}$ and project onto $T\Sigma$. This gives
\begin{equation}\label{eq:weyl-tangential-divergence}
 \bigl(\nu^c\gamma_{e(b}\lambda^{fg}\nabla_fW_{ca)}{}^e{}_g\bigr)^\top\overset{\Sigma}{=}d\bigl(\nu^c\nu^eW_{c(a}{}^f{}_e\gamma_{b)f}\bigr)^\top+\bigl(\nu^c\Pi_e^{\top f}\nabla_fW_{(b|c|}{}^e{}_{a)}\bigr)^\top.
\end{equation}
The last contribution is the tangential divergence
appearing in $\mathcal E_{ab}$. We must still simplify the contracted derivative
on the left-hand side of the above display. For this purpose, take the divergence of
Equation~\eqref{eq:U}. Two contributions from differentiating the right-hand
side of Equation~\eqref{eq:U} contain
$\nabla_g(W_{ae}{}^g{}_f\lambda^{ef})$
and the corresponding expression with $a$ replaced
by $b$. Both vanish by Equation~\eqref{eq:div-Wlambda},
\[
 \nabla_g(W_{ae}{}^g{}_f\lambda^{ef})=(d-2)C_{aef}\lambda^{ef}\overset{\Sigma}{=}0.
\]
The remaining pair gives
\[
 \tfrac1d\bigl(\nabla_b(W_{ae}{}^c{}_f\lambda^{ef})-\nabla_a(W_{be}{}^c{}_f\lambda^{ef})\bigr)=-\tfrac2d\nabla_{[a}(W_{b]e}{}^c{}_f\lambda^{ef}).
\]
Expanding the divergence of the left-hand side of
Equation~\eqref{eq:U} using
Equations~\eqref{pre:metrizability} and
\eqref{pre:contracted-bianchi} therefore yields
\begin{equation}\label{eq:weyl-full-divergence}
 \lambda^{ef}\nabla_eW_{ab}{}^c{}_f-(d+2)W_{ab}{}^c{}_f\nu^f+(d-2)C_{abf}\lambda^{fc}\overset{\Sigma}{=}-\tfrac2d\nabla_{[a}(W_{b]e}{}^c{}_f\lambda^{ef}).
\end{equation}

To evaluate the antisymmetrized derivative in the above, recall that
$H^{BC}=L(\lambda)^{BC}$ is the metrizability tractor
determined by the physical metric.
The mixed component of $\nabla_bH^{BC}$ is
$-W_{be}{}^c{}_f\lambda^{ef}/d$.
Its antisymmetrized derivative gives the mixed
component of $[\nabla_a,\nabla_b]H^{BC}$ on $\Sigma$,
since the additional contributions proportional
to $\nabla H$ vanish by Equation~\eqref{eq:normality}.
Computing this component from tractor curvature gives
\begin{equation}\label{eq:mixed-commutator}
 W_{ab}{}^c{}_e\nu^e-C_{abe}\lambda^{ec}\overset{\Sigma}{=}-\tfrac2d\nabla_{[a}(W_{b]e}{}^c{}_f\lambda^{ef}).
\end{equation}
Substituting Equation~\eqref{eq:mixed-commutator}
into Equation~\eqref{eq:weyl-full-divergence} yields
\[
 \lambda^{ef}\nabla_eW_{ab}{}^c{}_f\overset{\Sigma}{=}(d+3)W_{ab}{}^c{}_f\nu^f-(d-1)C_{abf}\lambda^{fc}.
\]
Contracting with $\nu$ and $\gamma$ and taking the
symmetric tangential part gives
\[
 \bigl(\nu^c\gamma_{e(b}\lambda^{fg}\nabla_fW_{ca)}{}^e{}_g\bigr)^\top\overset{\Sigma}{=}(d+3)\bigl(\nu^c\nu^eW_{c(a}{}^f{}_e\gamma_{b)f}\bigr)^\top-(d-1)(\nu^cC_{c(ab)})^\top.
\]
Substituting this expression into
Equation~\eqref{eq:weyl-tangential-divergence}
eliminates the contracted derivative.
Using Equation~\eqref{eq:curlye}, we obtain
\begin{equation}\label{eq:N-reduction}
 \bigl(\nu^c\nu^eW_{c(a}{}^f{}_e\gamma_{b)f}-\nu^cC_{c(ab)}\bigr)^\top\overset{\Sigma}{=}\tfrac13\mathcal E_{ab}.
\end{equation}
Since the trace term in Equation~\eqref{eq:IV-AB}
vanishes, we conclude
\[
 \IV_{ab}=\tfrac{2\alpha(d-7)}3\mathcal E_{ab}.
\]
\subsection{Matter constraints
(Section~\ref{sec:boundary-summary})}\label{app:matter-constraints}

We prove Equation~\eqref{eq:matter-boundary} and then
use the Einstein equations to derive the boundary
constraints in Equations~\eqref{eq:final-constraints}
and \eqref{eq:curvature-matter}.

The calculation in Equation~\eqref{eq:IV-AB}
depends on the first transverse derivative of the
three-tensor coefficient $\Omega$. For the source cotractor
in Equation~\eqref{eq:matter-tractor}, this coefficient
is $2\tau^{q+1}\widetilde\nabla_cS_{ab}$.
Equation~\eqref{eq:matter-change} and
$\nu^e\nabla_e\tau\overset{\Sigma}{=}2\alpha$ give
\[
 \nu^e\nabla_e(2\tau^{q+1}\widetilde\nabla_cS_{ab})=4q\alpha^2\tau^{q-1}\bigl(2(q+1)\mu_cS_{ab}+\mu_aS_{cb}+\mu_bS_{ac}\bigr)+O(\tau^q).
\]
The above vanishes after projection onto $T\Sigma$. The boundary calculation leading to
Equation~\eqref{eq:IV-AB} therefore applies to the
source cotractor $S_{CAB}$.
For $q=1$, the required normal contractions are
\[
 \bigl(\nu^c\nu^e\nabla_e(2\tau^{q+1}\widetilde\nabla_cS_{ab})\bigr)^\top\overset{\Sigma}{=}16\alpha^2S^\top_{ab},
\]
and
\[
 \bigl(\nu^c\nu^e\nabla_e(2\tau^{q+1}\widetilde\nabla_{(a}S_{b)c})\bigr)^\top\overset{\Sigma}{=}4\alpha^2S^\top_{ab}.
\]
Substituting these values into the boundary formula
in Equation~\eqref{eq:IV-AB} gives, for $q=1$,
\[
 \left.(\nu^c\Phi_{cba})^\top\right|_\Sigma=4\alpha^2(7-d)S^\top_{ab}+4\alpha^2\gamma_{ab}\lambda^{ef}S^\top_{ef}.
\]
For $q>1$, this contraction is zero.
Using $\tau_{ab}=2\alpha\gamma_{ab}$ proves
Equation~\eqref{eq:matter-boundary}.

For solutions of the Einstein equations, Equation~\eqref{eq:omega-matter}
identifies the source coefficient with $\Omega_{abc}$
and hence gives $\Phi_{cba}=\mathcal H_{cba}$.
The vanishing of the boundary value of the source coefficient
therefore implies $\III_{ab}=0$ by Equation~\eqref{eq:iii-definition}, while its computed
contraction gives $\IV_{ab}$.

Equations~\eqref{eq:trace-zero} and
\eqref{eq:matter-inverse} imply
\[
 (d-1)S^\top_{ab}\overset{\Sigma}{=}T^\top_{ab},\qquad \lambda^{ab}S^\top_{ab}\overset{\Sigma}{=}0.
\]
Consequently, the trace contribution in
Equation~\eqref{eq:matter-boundary} vanishes.
For $q=1$, the remaining contribution is
$4\alpha^2(7-d)T^\top_{ab}/(d-1)$; for $q>1$ it is zero.
With $\alpha|_\Sigma=-1$, these are precisely the
fourth-form constraints in
Equation~\eqref{eq:final-constraints}.

To derive the curvature--matter relation in every
dimension, we use Equation~\eqref{eq:K-curvature},
which relates~$\mathcal E_{ab}$ directly to the
difference of the two normal contractions of $Q$.
On solutions of Equation~\eqref{eq:efe},
Equation~\eqref{eq:omega-matter} identifies these with
the source derivatives computed above. Their difference
therefore gives
\[
 \bigl(\nu^c(Q_{abc}-Q_{c(ab)})\bigr)^\top\overset{\Sigma}{=}
 \begin{cases}
  \tfrac{12\alpha^2}{d-1}T^\top_{ab}, & q=1,\\[3pt]
  0, & q>1.
 \end{cases}
\]
Comparing with Equation~\eqref{eq:K-curvature} yields
$\mathcal E_{ab}=-6\alpha T^\top_{ab}/(d-1)$ for $q=1$
and $\mathcal E_{ab}=0$ for $q>1$.
Using $\alpha|_\Sigma=-1$ proves
Equation~\eqref{eq:curvature-matter}. In particular, for $q=1$, dimension seven gives
$\mathcal E_{ab}=T^\top_{ab}$ while~$\IV_{ab}=0$.

\end{document}